\documentclass[10pt]{emulateapj}
\usepackage{apjfonts}  
\usepackage{graphicx}  
\usepackage{psfig}  
  
\slugcomment{{\sc Submitted to ApJ: January 26, 2006} }  
\catcode`\@=11  
\def\@versim#1#2{\vcenter{\offinterlineskip  
        \ialign{$\m@th#1\hfil##\hfil$\crcr#2\crcr\sim\crcr } }}  
\newcommand{\begit}{\begin{itemize}}  
\newcommand{\enit}{\end{itemize}}  
\newcommand{\begen}{\begin{enumerate}}  
\newcommand{\enen}{\end{enumerate}}

\newcommand{\beq}{\begin{equation}}  
\newcommand{\eeq}{\end{equation}}

\newcommand{\nughz}{\nu_{\rm GHz}}

\newcommand{\beqa}{\begin{eqnarray}}   
\newcommand{\eeqa}{\end{eqnarray}}   
  
\newcommand{\sds}{\dot{\Sigma}_\star}

\def\lsim{\mathrel{\mathpalette\@versim<}}  
\def\gsim{\mathrel{\mathpalette\@versim>}}

\begin{document}  
  
\title{Magnetic Fields in Starburst Galaxies {\small and} The Origin
of the FIR-Radio Correlation}
  
\author{Todd A.~Thompson,\altaffilmark{1,2}   
Eliot Quataert,\altaffilmark{3}  
Eli Waxman,\altaffilmark{4}  
Norman Murray,\altaffilmark{5,6} \&  
Crystal L.~Martin\altaffilmark{7}}  
  
\altaffiltext{1}{  
Department of Astrophysical Sciences, Peyton Hall-Ivy Lane,  
Princeton University, Princeton, NJ 08544; thomp@astro.princeton.edu}  
\altaffiltext{2}{Lyman Spitzer Jr.~Fellow}  
\altaffiltext{3}{Astronomy Department   
\& Theoretical Astrophysics Center, 601 Campbell Hall,   
The University of California, Berkeley, CA 94720;   
eliot@astro.berkeley.edu}  
\altaffiltext{4}{Physics Faculty, Weizmann Institute of Science, Rehovot 76100, Israel;  
waxman@wicc.weizmann.ac.il}  
\altaffiltext{5}{Canada Research Chair in Astrophysics}  
\altaffiltext{6}{Canadian Institute for Theoretical Astrophysics,   
60 St.~George Street, University of Toronto, Toronto,  
ON M5S 3H8, Canada; murray@cita.utoronto.ca}  
\altaffiltext{7}{Department of Physics, The  
University of California, Santa Barbara, CA 93106;   
cmartin@physics.ucsb.edu}

\begin{abstract}  
  
We estimate minimum energy magnetic fields ($B_{\rm min}$) for a  
sample of galaxies with measured gas surface densities. The sample  
spans more than four orders of magnitude in surface density from  
normal spirals to luminous starbursts.  We show that the ratio of the  
minimum energy magnetic pressure to the total pressure in the ISM  
decreases substantially with increasing surface density.  For the  
ultra-luminous infrared galaxy Arp 220, this ratio is $\sim10^{-4}$.  
Therefore, if the minimum energy estimate is applicable, magnetic  
fields in starbursts are dynamically weak compared to gravity, in  
contrast to our Galaxy and other normal star-forming spiral galaxies.  
We argue, however, that rapid cooling of relativistic electrons in  
starbursts invalidates the minimum energy estimate.  We critically  
assess a number of independent constraints on the magnetic field  
strength in starburst galaxies.  In particular, we argue that the  
existence of the FIR-radio correlation implies that 
the synchrotron cooling timescale for cosmic ray electrons is much
shorter than their escape time from the galactic disk; this in turn
implies that the true magnetic field in starbursts is significantly 
larger than $B_{\rm min}$.  The strongest argument {\it against} such 
large fields is that one might expect starbursts to have steep radio 
spectra indicative of strong synchrotron cooling, which is not observed.  
We show, however, that ionization and bremsstrahlung losses can flatten 
the nonthermal spectra of starburst galaxies even in the presence of rapid 
cooling, providing much better agreement with observed spectra.  We further
demonstrate that ionization and bremsstrahlung losses are likely to be
important in shaping the radio spectra of most starbursts at GHz
frequencies, thereby preserving the linearity of the FIR-radio
correlation.  We thus conclude that magnetic fields in starbursts are
significantly larger than $B_{\rm min}$.  We highlight several
observations that can test this conclusion.
\end{abstract}  
  
  
\section{Introduction}  
\label{section:intro}  
  
The magnetic energy density of the Galaxy is observed to be in rough  
equipartition with the cosmic ray energy density and the turbulent  
pressure. The sum of these yields a total midplane pressure consistent  
with that required by hydrostatic equilibrium, given the mass  
distribution in the solar neighborhood (Boulares \& Cox 1990). This  
equipartition magnetic field is roughly 6 $\mu$G at the solar circle  
(e.g., Beck 2001). Similar magnetic field strengths are found in most  
normal star-forming spiral galaxies (Fitt \& Alexander 1993; Niklas  
1995).  
  
\begin{figure*}  
\centerline{\hbox{\psfig{file=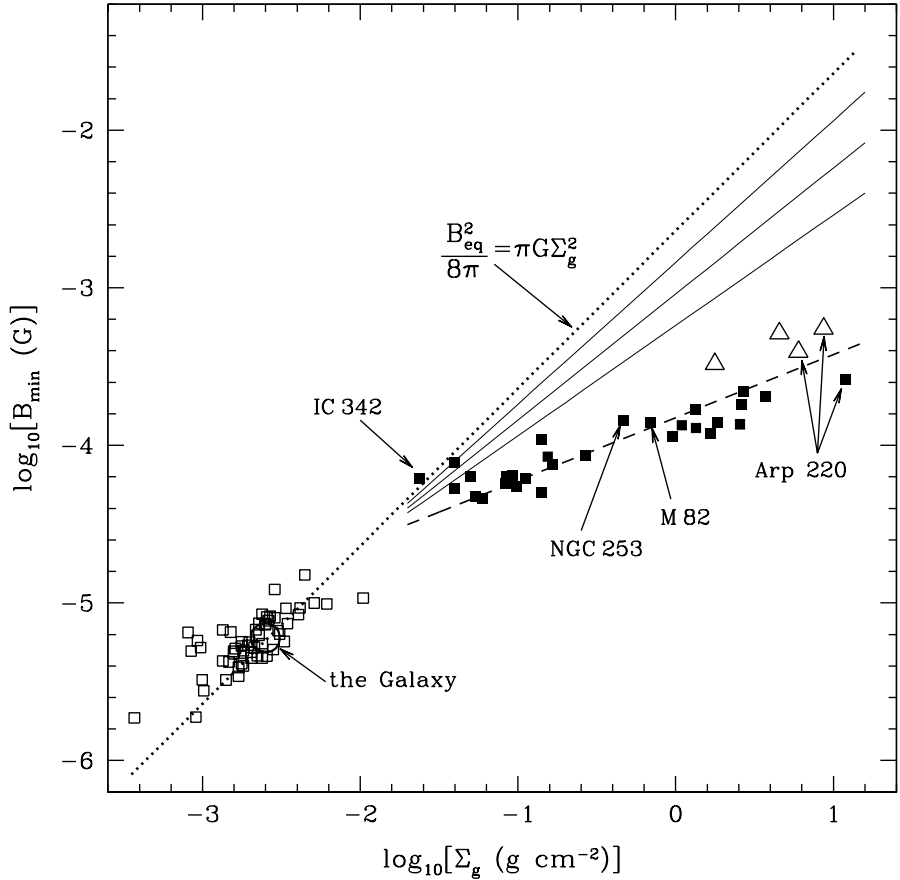,width=12.5cm}}}
\figcaption[x]{Minimum energy magnetic field ($B_{\rm min}$;
eq.~[\ref{bmin}]) versus measured gas surface density ($\Sigma_g$).
Normal star-forming galaxies (open squares; Table \ref{table:spiral}),
starburst galaxies (filled squares; Table \ref{table:starburst}), and
the Galaxy ($B_{\rm min}=6$ $\mu$G at the solar circle, e.g., Beck
2001; $\Sigma_g\simeq2.5\times10^{-3}$ g cm$^{-2}$, Boulares \& Cox
1990) are shown.  To indicate the uncertainties at high $\Sigma_g$,
the open triangles show $B_{\rm min}$ for IC 883 (Arp 193), Mrk 273,
and the individual nuclei of Arp 220, as inferred using $\Sigma_g$ and
the radial size from Downes \& Solomon (1998) (their ``extreme''
starbursts; see their Table 12) and with the radio flux from Condon et
al.~(1991) at 8.44 GHz (see Table \ref{table:extreme}).  The dotted
line is the equipartition magnetic field (eq.~[\ref{bequipartition}])
and the dashed line is the scaling $B_{\rm min} \propto
\Sigma_g^{2/5}$ (eq.~[\ref{bminfir}]).  The latter scaling is expected
if the cosmic ray electron cooling timescale is much shorter than the
escape timescale from the galactic disk, in which case the true
magnetic field is also significantly larger than the minimum energy
estimate (see \S\ref{section:for}; eq.~[\ref{bminfir}]).  The solid
lines show the scalings $B \propto \Sigma_g^a$ (with a = 0.9, 0.8, and
0.7) used in \S \ref{section:for_breaks}. \\
\label{fig:bmin}}  
\end{figure*}

Starbursts and ultra-luminous infrared galaxies (ULIRGs) have much  
higher surface densities and turbulent velocities than local spirals.  
As an example, Arp 220 has a gas surface density $\sim 10^4$ times  
higher than the Galaxy (Downes \& Solomon 1998).  If Arp 220 has a  
magnetic energy density in rough equipartition with its hydrostatic  
pressure, the implied magnetic field strength would be   
$\sim0.03$\,G on few-hundred parsec scales (Fig.~\ref{fig:bmin}, 
Table \ref{table:starburst}).  This is substantially  
larger than the value of $\sim 1$ mG typically inferred in  
ULIRGs (e.g., Condon et al.~1991) using the observed radio emission  
and the classic ``minimum energy'' argument (Burbidge 1956; Longair  
1994).    
  
Throughout this paper we use the term ``minimum energy'' to  
refer to the magnetic field strength inferred using the observed radio  
flux and assuming comparable cosmic ray and magnetic energy densities  
($B_{\rm min}$; eq.~[\ref{bmin}]).  We reserve the term ``equipartition'' for magnetic  
field strengths that are dynamically important with respect to gravity  
($B_{\rm eq}$; eq.~[\ref{bequipartition}]).  
  
We find that the large discrepancy between the minimum energy and
equipartition magnetic field estimates in Arp 220 is generic to
starburst galaxies; this discrepancy motivates the analysis of
magnetic field strengths in luminous starbursts presented in this
paper. In \S\ref{section:results} we present a sample of galaxies with
measured gas surface densities and radio fluxes, for which both
minimum energy and equipartition field strengths can be estimated.  In
\S\ref{section:against} we summarize arguments in favor of $B \sim
B_{\rm min}$ and against significantly larger fields.  Rebuttals to a
subset of these arguments are given in \S\ref{section:for}, where we
also provide independent arguments why magnetic fields in starbursts
are likely to be $\gg B_{\rm min}$.  
These arguments draw heavily on the FIR-radio correlation for
star-forming galaxies and the observed radio spectra of starbursts.
Finally, in \S\ref{section:discussion} we summarize our results and
discuss their implications.
  
\

\
  
\section{Data \& Results}  
\label{section:results}  
  
Longair (1994) derives the minimum energy magnetic field strength as  
(Vol.~II, eq.~[19.30])  
\beq B_{\rm min}\approx7\times10^{-4}\delta_{2}^{2/7}L_{\nu,  
\,23}^{2/7}\, V_{6}^{-2/7}\,\nu_{\rm GHz}^{1/7}\,\,{\rm G}  
\label{bmin}  
\eeq where $L_{\nu, \,23}=L_\nu/10^{23}$ W Hz$^{-1}$, $\nu_{\rm
GHz}=\nu/10^9$ Hz, $V_{6}=V/(100)^3$ pc$^3$ is the volume of the
emitting region, $\delta_{2}=\delta/10^2$ is the ratio of the energy
in cosmic ray ions to the energy in cosmic ray electrons, and the
derivation assumes that the cosmic ray electron energy spectrum is
$n(\gamma) \propto \gamma^{-p}$ with $p = 2.5$.  We have scaled the
results above for parameters appropriate to compact starbursts.  For
parameters typical of the Galaxy, $L_{\nu, \,23}\sim3\times10^{-2}$
and $V_6\sim7\times10^4$ so that $B_{\rm min}\approx10^{-5}$ G.
  
The minimum energy magnetic field strength is often referred to as  
``equipartition'' because the corresponding magnetic energy density  
($U_{B_{\rm min}}$) is approximately equal to the total cosmic ray  
(electron + ion) energy density if $B_{\rm min}$ actually obtains.  In  
this paper, however, we use the term ``equipartition'' to denote  
magnetic energy densities comparable to the total hydrostatic pressure  
of the ISM.  For a gas disk of surface density $\Sigma_g$, hydrostatic  
equilibrium with the self-gravity of the disk implies a midplane  
pressure \beq P\approx\pi G \Sigma_g^2.  
\label{gaspressure}   
\eeq We measure equipartition via the parameter $\eta$ defined by  
$U_B=\eta P$.  The equipartition ($\eta=1$) field strength is thus  
\beq B_{\rm eq}\approx(8\pi^2  
G)^{1/2}\Sigma_g\approx 2 \, \Sigma_g \,\,{\rm mG},  
\label{bequipartition}  
\eeq where $\Sigma_g$ is in cgs units.  Theoretically, there are
significant uncertainties about the origin of magnetic fields in
galactic disks, and so it is difficult to estimate the expected value
of $\eta$.  Field strengths as large as equipartition are, however,
possible if the magnetic energy density equilibrates with the
turbulent energy density of the ISM.  The scale-height $h$ of a
self-gravitating galactic disk is given by $h \sim \delta v^2/2 \pi G
\Sigma_g$, where $\delta v$ is the turbulent velocity.  As a result,
the energy density in turbulent motions is $\rho \,\delta v^2\sim\pi
G\Sigma_g^2$, where $\rho$ is the mean density of the ISM.  Thus, if
the field is amplified by the turbulent motions so that
$B^2/8\pi\sim\rho\delta v^2$, equipartition follows.
  
The gravitational potential of normal star-forming galaxies includes a  
significant contribution from stars (and perhaps dark matter) on the  
scale of their effective radii.  Neglecting for simplicity the  
different scale-heights of stars and gas, equation (\ref{gaspressure})  
should then be replaced by $P\approx\pi G \Sigma_g\Sigma_{\rm  
tot}\approx\pi G \Sigma_g^2 f_g^{-1}$, where $f_g$ is the gas fraction  
and $\Sigma_{\rm tot}$ is the total surface density.  Typically, $f_g$  
is in the range $\sim0.1-0.2$ for normal galaxies and $f_g\sim0.5$ for  
starbursts.  Thus, the magnetic field could in principle be larger  
than equation (\ref{bequipartition}) by a factor of $f_g^{-1/2}$ and  
still be sub-dominant with respect to gravity.  In this paper, these  
distinctions are not important because we focus on gas-rich starbursts  
and use equation (\ref{bequipartition}) primarily as a useful point of  
comparison for field strengths inferred using the minimum energy  
argument (eq.~[\ref{bmin}]).  
  
Figure \ref{fig:bmin} shows the minimum energy magnetic field $B_{\rm  
min}$ inferred using equation (\ref{bmin}) as a function of the  
surface density $\Sigma_g$ of the galactic disk for a sample of  
galaxies ranging from local spirals to ULIRGs.  The sample is based on  
galaxies with surface densities and sizes used by Kennicutt (1998) to  
study the global properties of star formation across a range of  
galaxies (the ``Schmidt Laws'').  The properties of the normal  
star-forming and starburst galaxies are listed in Tables \ref{table:spiral}  
and \ref{table:starburst}, respectively.  
Systems described as starbursts by Kennicutt are  
denoted here by solid squares, whereas his normal star-forming  
galaxies are labeled by open squares.   
  
In computing the radio emitting volume (eq.~[\ref{bmin}]), we assume that the scale height   
of the synchrotron emission is $h_{rad}=500$\,pc for the normal galaxies (Beuermann et al.~1985;  
Dumke \& Krause 1998) and $h_{rad}=100$\,pc for the starbursts (filled squares)  
(e.g., Klein et al.~1988; Seaquist \& Odegard 1991;   
Condon et al.~1991).\footnote{$B_{\rm min} \propto h_{rad}^{-2/7}$ so  
uncertainties in $h_{rad}$ introduce only modest uncertainties in  
$B_{\rm min}$.}  We use radial sizes from Kennicutt (1998); for normal  
galaxies, these are taken from the RC2 catalog (de Vaucouleurs et  
al.~1976).  Detailed studies show that for spiral galaxies the ratio  
of the optical to radio disk scale length is $\sim0.9-1.3$   
(Hummel 1980; Fitt \& Alexander 1993).    
For both the normal galaxies (open squares) and the starbursts (filled squares),   
we have computed $B_{\rm min}$ using the radio luminosity at 1.4 GHz.  
  
The open triangles illustrate the uncertainties in the results of
Figure \ref{fig:bmin} at high $\Sigma_g$.  These points denote the
``extreme'' starbursts of Downes \& Solomon (1998) (see their Table
12): IC 883 (Arp 193), Mrk 273, and the two individual nuclei of Arp
220.  For these systems, we use the gas surface densities and radial
sizes from Downes \& Solomon (instead of Kennicutt 1998), and we
combine this data with radio fluxes from the 8.44\,GHz observations of
Condon et al.~(1991), which have sufficient resolution to probe the
same spatial scales as the observations of Downes \& Solomon (in
contrast to the 1.4\,GHz observations used for the rest of the systems
in Fig.~\ref{fig:bmin}).
The properties of these starbursts are collected in Table
\ref{table:extreme}.  Although the inferred $B_{\rm min}$ is somewhat
larger at a given $\Sigma_g$ than for the starburst sample in Table
\ref{table:starburst} (filled squares in Fig. \ref{fig:bmin}), the
reasonably close correspondence is encouraging.
  
In addition to the inferred minimum energy magnetic field, Figure
\ref{fig:bmin} also includes several theoretical curves.  The dotted
line shows the equipartition magnetic field
(eq.~[\ref{bequipartition}]) while the dashed line shows the scaling
$B_{\rm min} \propto \Sigma_g^{2/5}$ derived in
\S\ref{section:for_fig1}.  Finally, the solid lines show the scalings
$B \propto \Sigma_g^a$ (with a = 0.9, 0.8, and 0.7) used in our
discussion of the FIR-radio correlation in \S\ref{section:for_breaks}.
Although these alternative (non-equipartition) scalings are somewhat
arbitrary, the particular choice $B \propto \Sigma_g^{0.7}$ is physically
motivated by setting the magnetic energy density equal to the 
pressure in the ISM produced by star formation ($P_\star$). Because
$P_\star\propto\sds$ (where $\sds$ is the star formation rate per unit
area; see, e.g., Chevalier \& Clegg 1985, Thompson, Quataert, \&
Murray 2005) and because the Schmidt law for star formation is
$\sds\propto\Sigma_g^{1.4}$ (Kennicutt 1998), the scaling $B \propto
\Sigma_g^{0.7}$ follows.

Our estimates of $B_{\rm min}$ in Figure \ref{fig:bmin} for normal star-forming  
galaxies are in good agreement with those estimated by other authors  
(e.g., Fitt \& Alexander 1993; Niklas 1995; Beck 2000).   
When these minimum energy fields can be compared with other magnetic field estimates  
(e.g., from Faraday rotation) they are generally found to be consistent  
(e.g., Beck 2000; Valle\'e 1995).  In addition, the crucial  
assumption of the minimum energy argument, namely that the energy  
density in the magnetic field is roughly equal to the energy density  
in cosmic rays, is confirmed locally in the Galaxy using $\gamma$-ray  
observations (Strong et al.~2000).  
  
In keeping with previous studies, our results in Figure \ref{fig:bmin}
show that in normal star-forming galaxies, $B_{\rm min} \sim B_{\rm
eq}$.  However, the striking result from Figure \ref{fig:bmin} is that
the starburst galaxies systematically have $B_{\rm min} < B_{\rm eq}$.
The ratio $\eta_{\rm min}=(B_{\rm min}^2/8\pi)/P$ is a strongly
decreasing function of $\Sigma_g$; for a typical starburst such as
M82, $\eta_{\rm min} \sim10^{-2}$, while for Arp 220, $\eta_{\rm min}
\sim 10^{-4}$.\footnote{Free-free absorption may suppress the observed
radio flux density at GHz frequencies in dense starbursts (e.g.,
Condon et al. 1991). However, even in our most extreme case, the ULIRG
Arp 220, this correction amounts to a small increase ($\sim10$\%) in
our inferred $B_{\rm min}$.  Another uncertainty is the contribution
of free-free emission to the observed radio flux.  At GHz frequencies
the thermal fraction is of order 10\% for normal star-forming galaxies
(e.g., Niklas 1995).  For starbursts the thermal fraction is less
certain.  In any case, subtracting the thermal contribution to the
radio luminosity would decrease the inferred $B_{\rm min}$,
strengthening our conclusion that $B_{\rm min}\ll B_{\rm eq}$ in
starbursts.}
  
If the minimum energy magnetic field strength estimate is applicable,
the direct implication of Figure \ref{fig:bmin} is that in starbursts,
magnetic fields are dynamically weak compared to gravity in the phase
of the ISM in which the radio emitting electrons reside.  Because of
the potential importance of this conclusion, it is worth critically
examining the applicability of the minimum energy estimate in the
context of luminous starbursts (see Beck \& Krause 2005 for a more
general discussion). The minimum energy argument rests on the
assumption that the magnetic energy density is comparable to the
cosmic ray energy density.  Given that the magnetic field and cosmic
rays could have rather different energy generation and dissipation
mechanisms (``sources and sinks''), the generic applicability of this
assumption is unclear.
  
In the context of local spirals, one argument in favor of  
the minimum energy estimate is precisely that it implies a total  
cosmic ray + magnetic energy density in approximate equipartition with  
gravity.  If $B \ne B_{\rm min}$, the net cosmic ray + magnetic energy  
density would exceed the gravitational binding energy of the gas, and  
the field and relativistic particles would presumably escape from the  
galactic disk (e.g., Parker 1965, 1966; Duric 1990).  This provides a natural mechanism  
by which the magnetic field strength can adjust to a value of $\sim  
B_{\rm min}$.  However, such an argument does not apply in starbursts  
given that the minimum energy argument itself implies $\eta_{\rm min} \ll 1$  
(Fig. \ref{fig:bmin}).    
  
A more significant worry concerning the applicability   
of the minimum energy argument in starbursts is that strong synchrotron and   
inverse Compton (IC) losses could lead to rapid electron cooling.  If the steady   
state electron energy density is significantly lower as a result, then the minimum   
energy magnetic field is an {\it underestimate}.    
  
In the following sections we critically review independent  
observational and theoretical constraints on the magnetic fields in  
starbursts.  We begin by summarizing arguments in favor of $B \sim B_{\rm min}$ and
against the hypothesis that magnetic fields in starbursts are
significantly larger.  Several of these arguments draw heavily on
existing results in the literature, but are worth reviewing in the
present context.  Rebuttals to these arguments are given in
\S\ref{section:for}, where we also provide independent arguments in
favor of $B\gg B_{\rm min}$.
  
\section{Arguments For $B\sim B_{\lowercase {\rm min}}$ in Starburst Galaxies}  
  
\label{section:against}  
  
\subsection{Zeeman Measurements}  
\label{section:against_zeeman}  
  
To our knowledge, there are no extragalactic detections of Zeeman
splitting.  In four ULIRGs, however, Killeen et al.~(1996) derive
upper limits of $B\lesssim 3-5$ mG using OH masers.  These limits are
larger than the minimum energy estimate of $\sim$1\,mG for ULIRGs, but
are a factor of $\sim$5 smaller than $B_{\rm eq}$ for Arp 220 in
Figure \ref{fig:bmin}.  In our own Galaxy OH masers trace regions of
higher than average surface density and magnetic field strength, with
$B \approx 2-10$ mG (e.g., Fish et al.~2003); it is unlikely that the
opposite is true in ULIRGs, in which case the Killeen et al.~(1996)
upper limits argue against $B \sim B_{\rm eq}$.  However, a larger
sample of Zeeman measurements is needed, particularly for galaxies
which also have reliable gas surface density measurements.
  
\subsection{Cooling Breaks}  
\label{section:against_breaks}  
  
The synchrotron cooling timescale for cosmic ray electrons emitting at frequency  
$\nu$ is   
\beq   
\tau_{\rm syn}\approx10^{6}\,\,B_{100}^{-3/2}\,\nughz^{-1/2}\,\,{\rm yr},   
\label{syncool}  
\eeq   
where $B_{100} = B/100 \mu G$.  IC losses can be appreciable as  
well and lead to cooling on a timescale   
\beq   
\tau_{\rm IC} \approx 5 \times 10^{5}\,\,B_{100}^{1/2}\,\nughz^{-1/2}\,U^{-1}_{\rm ph,-9}\,{\rm yr},   
\label{iccool}  
\eeq   
where $U_{\rm ph} = 10^{-9} U_{\rm ph,-9}$ ergs  
cm$^{-3}$ is the energy density of (primarily infrared) photons in the  
starburst.   
  
The cooling timescale for electrons can be indirectly constrained using the  
radio spectrum.  For continuous injection of relativistic  
electrons, if the cooling time   
\beq   
\tau_{\rm cool}^{-1} = \tau_{\rm syn}^{-1} + \tau_{\rm IC}^{-1}   
\eeq   
of the electrons emitting at  
frequency $\nu$ is shorter than the escape time from the galactic  
disk, the radio spectrum will steepen by $\Delta \alpha = 1/2$ at higher frequencies,   
where $F_\nu \propto \nu^{-\alpha}$.  For canonical electron power-law  
indices of $p = 2$ expected theoretically for strong shocks (e.g.,  
Blandford \& Eichler 1987) and observed {\it in situ} in some  
supernova remnants (e.g., Aharonian et al.~2005; Brogan et al.~2005),  
$\alpha = 1/2$ in the absence of cooling and steepens to $\alpha = 1$  
in the presence of cooling.  The escape time of relativistic particles  
from the galactic disk is uncertain.  In local spirals the escape is  
due to diffusion across the magnetic field, while in starbursts it may  
be much more rapid due to advection out of the galaxy with a galactic  
wind.  The escape time in the latter case is roughly   
\beq  
\tau_{\rm esc}\sim h/v_w \sim 3 \times 10^5 \, h_{100} \, v_{500}^{-1}\,\,\,{\rm  yr},   
\label{tau_esc}  
\eeq  
where $h = 100 \, h_{100}$ pc is the scale height of the galactic disk and  
$v_w = 500 \, v_{500}$ km s$^{-1}$ is the speed of the galactic wind  
(see, e.g., Martin 1999 for estimates of the latter).  
  
The radio spectra of both normal spirals and luminous starbursts are  
similar with $\langle \alpha \rangle \approx 0.75$ from $\approx 1-10$  
GHz (e.g., Condon et al. 1991; Condon 1992; Niklas et al. 1997).  
There appears to be no strong variation of $\alpha$ with the  
luminosity of the galaxy (though Niklas et al.~1997 note a weak trend  
for starbursts to have somewhat flatter nonthermal spectral indices).  
The lack of a systematic variation of the spectral index with  
luminosity, together with the fact that the observed radio spectra are  
significantly flatter than expected for a ``cooled'' electron  
distribution, suggests that $\tau_{\rm cool} \gtrsim \tau_{\rm esc}$,  
even in luminous starbursts.  Using equation (\ref{tau_esc}) to  
estimate $\tau_{\rm esc}$, equation (\ref{syncool}) then implies field  
strengths $\lesssim 2 \times 10^{-4}$ G, reasonably consistent with  
the minimum energy estimates for the starbursts in Figure  
\ref{fig:bmin} and {inconsistent} with $B\gg B_{\rm min}$.  
  
\subsection{Synchrotron Halos}  
\label{section:against_halos}  
  
The spatial variation of the radio spectrum can provide a more  
stringent constraint on the importance of electron cooling than the  
integrated radio spectrum alone.  To take a concrete example,  
observations of M82 by Klein et al.~(1988) and Seaquist \& Odegard  
(1991) reveal an extended radio halo along the minor axis of M82,  
coincident with the observed galactic wind.  The radio spectrum at GHz  
frequencies is observed to be roughly constant in the disk of M82, but  
to steepen significantly in the halo a distance $h_{\rm break} \sim  
300$ pc above the midplane.  The usual interpretation of this halo is  
that relativistic electrons generated in the starburst are advected  
out with the galactic wind.  The steepening at $h_{\rm break} \sim  
300$ pc can then be modeled as a consequence of electron cooling  
(e.g., Seaquist \& Odegard 1991).  The radio emitting electrons in the  
halo must thus have $h_{\rm break} \sim \tau_{\rm cool} v_w$.  This  
approximate equality can be used to put a limit on the magnetic field  
in the disk of M82.  For example, the equipartition field of $B_{\rm  
eq} \approx 1.6$ mG in M82 (see Fig.~\ref{fig:bmin}) implies a  
synchrotron cooling time of $\tau_{\rm syn} \sim 10^4$ yrs, which, in  
turn, requires an unphysically large $v_w \sim 30,000$ km s$^{-1}$ to  
explain the observed scale of the spectral break in M82.  Thus a  
magnetic field strength of $\sim B_{\rm eq}$ is strongly disfavored  
under the interpretation that the synchrotron halo in M82 is due to  
electrons accelerated in the disk and advected out in a galactic wind.  
By contrast, a field strength of $B \sim B_{\rm min}$ implies $v_w  
\sim 500$ km s$^{-1}$ to explain a spectral break at $h_{\rm  
break}\sim 300$ pc --- much more reasonable in the context of galactic  
winds. Thus, the spatial variation of the radio spectrum in M82  
appears to provide independent evidence that $\tau_{\rm  
cool}\gtrsim\tau_{\rm esc}$ in the galactic disk and hence that $B\sim  
B_{\rm min}\ll B_{\rm eq}$.  
  
In fact, more careful estimates account for both IC cooling and  
adiabatic cooling of relativistic electrons as they are advected out  
into the halo.  This increases the required wind speed by a factor of  
few to $\sim2000$ km s$^{-1}$ (e.g., Seaquist \& Odegard 1991).  Although this  
is substantially smaller than the speed of $\sim30,000$ km s$^{-1}$  
required for $B\sim B_{\rm eq}$, it is larger than the inferred speeds  
of galactic winds (e.g., Martin 1999) and is comparable to the initial  
velocity of supernova ejecta.  One cannot, however, rule out a very  
low density phase of a galactic wind with the required velocity $\sim  
2000$ km s$^{-1}$.    
  
We note that NGC 253 (Carilli et al.~1992), NGC 1569 (Israel \& de  
Bruyn 1988; Lisenfeld et al.~2004), NGC 4631 (Ekers \& Sancisi 1977),  
NGC 891 (Allen, Sancisi, \& Baldwin 1978), and NGC 4666 (Dahlem et  
al.~1997) all exhibit radio halos qualitatively similar to those in  
M82.  Unfortunately, with the exception of NGC 253, these systems have  
lower gas surface densities than M82 and lie in the part of Figure  
\ref{fig:bmin} where $B_{\rm min} \sim B_{\rm eq}$.  Clearly, a more  
extensive survey for low surface brightness extended radio emission  
around starbursts with $\Sigma_g \sim 1$ g cm$^{-2}$ would be very  
useful in discriminating between the minimum energy and equipartition  
magnetic field strengths (see Table \ref{table:starburst}).    

\section{Arguments For $B\gg B_{\lowercase {\rm min}}$ in Starburst Galaxies}  
\label{section:for}  
  
In this section we rebut the arguments for $B\sim B_{\rm min}$
presented in Sections \ref{section:against_breaks} and
\ref{section:against_halos} and provide additional evidence that $B\gg
B_{\rm min}$.  Most of this section centers on showing that $\tau_{\rm
cool} \lesssim \tau_{\rm esc}$ is consistent with radio observations of
starbursts.
  
\subsection{Understanding $B_{\rm min}$ as a Function of $\Sigma_g$}  
\label{section:for_fig1}  
  
A magnetic field strength of $B \gg B_{\rm min}$ in starbursts would
imply that the cooling time for relativistic electrons is very short
compared to the escape time ($\tau_{\rm cool}\ll\tau_{\rm esc}$).  To
show this, we assume that $B=\eta^{1/2}B_{\rm eq}$, that $\tau_{\rm
esc} = 10^6 \tau_{\rm esc,6}$ yr, and we combine equations
(\ref{bequipartition}) and (\ref{syncool}) to estimate that \beq
\frac{\tau_{\rm syn}}{\tau_{\rm esc}} \sim
0.01\,\,\Sigma_g^{-3/2}\eta^{-3/4} \nu_{\rm GHz}^{-1/2} \tau_{\rm
esc,6}^{-1}. \label{taus} \eeq For $\Sigma_g \gtrsim 0.05 \, \tau_{\rm
esc,6}^{-2/3} \, \eta^{-1/2}$ g cm$^{-2}$, equation (\ref{taus})
implies that $\tau_{\rm syn}<\tau_{\rm esc}$ for electrons emitting at
GHz frequencies. Including IC cooling decreases this critical surface
density by a factor of order unity.  In addition, for lower surface
density galaxies, the electron escape time becomes increasingly
determined by diffusion across the magnetic field rather than escape
in a galactic wind.  This likely increases $\tau_{\rm esc}$
substantially and thus decreases the critical surface density where
$\tau_{\rm syn} \sim \tau_{\rm esc}$.
  
The above considerations show that if $B \gg B_{\rm min}$ then nearly
all of the starbursts in Figure \ref{fig:bmin} are in the limit where
$\tau_{\rm cool}<\tau_{\rm esc}$.  Our observationally inferred
results for $B_{\rm min}(\Sigma_g)$ can be understood quantitatively
in this limit as follows.  In the rapidly cooling limit, the electrons
radiate all of the energy supplied to them by supernova shocks and
thus the radio flux per unit volume is determined by the supernova
rate per unit volume: \beq \nu L_\nu/V \propto \sds/h,
\label{fir_radio}  
\eeq   
where $\sds$ is the star formation rate per unit area and $V$ is  
the radio emitting volume.  Using the Schmidt Law for star formation  
($\sds\propto \Sigma_g^{7/5}$; Kennicutt 1998) and equation  
(\ref{bmin}), we find \beq B_{\rm min} \propto  
\Sigma_g^{2/5}h^{-2/7}. \label{bminscale}\eeq   
To estimate the  
proportionality constant in equation (\ref{bminscale}), we assume that  
$10^{-3} \, \xi$ of each SN's energy is supplied to relativistic  
electrons (and is radiated away), in which case  
equation (\ref{bminscale}) becomes  
\beq   
B_{\rm min} \approx 7\times10^{-5}\delta_2^{2/7} \xi^{2/7}   
\Sigma_g^{2/5} h_{100}^{-2/7}\nu_{\rm GHz}^{-1/7}\,\,\,{\rm G},   
\label{bminfir}   
\eeq  
Taking $h \sim {\rm constant}$ and $\xi = 10$ (see \S\ref{section:for_fir}), 
we plot equation (\ref{bminfir}) in Figure \ref{fig:bmin} (dashed  
line) and find that it is in reasonable agreement with the results for  
starbursts.  This provides strong support for the hypothesis that  
$\tau_{\rm cool} \lesssim \tau_{\rm esc}$ in these systems.  
Furthermore, the agreement between the starburst data and equation  
(\ref{bminfir}) emphasizes that the inferred minimum energy field in  
starbursts is not a reliable magnetic field measurement, but rather  
simply reflects the fact that the radio flux is proportional to the  
star formation rate (eq.~[\ref{fir_radio}]).

\begin{figure*}  
\centerline{\hbox{\psfig{file=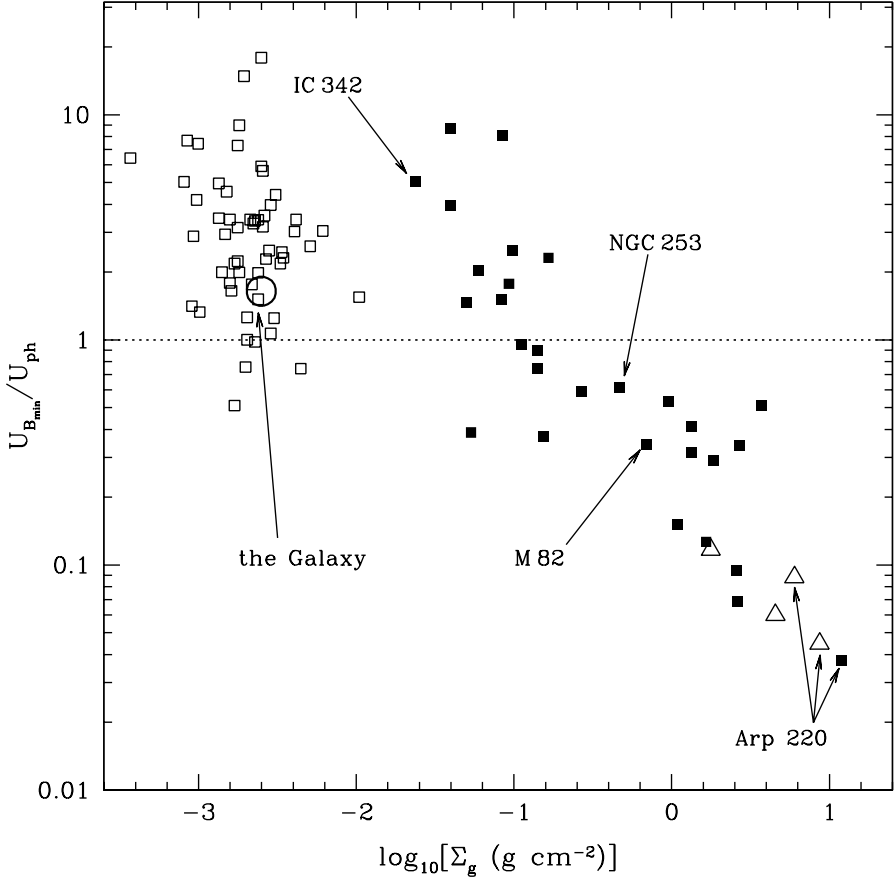,width=12.5cm}}}  
\figcaption[ubup]{The ratio of the minimum energy magnetic energy density to  
the photon energy density ($U_{B_{\rm min}}/U_{ph}$) as a function of  
the gas surface density ($\Sigma_g$). The symbols are the same as  
those in Fig.~\ref{fig:bmin}.  $U_{ph}$ is calculated using  
eq.~(\ref{uph}) (see also eq.~[\ref{uphtau}]) for all systems except the Galaxy, for which we take  
$F=L/\pi R^2$ with $L=1.5\times 10^{10}$ L$_\odot$ and $R=8.5$ kpc (as  
appropriate for the estimate of $B_{\rm min}$ at the solar circle).  
$U_{B} < U_{ph}$ is difficult to reconcile with the  
linearity of the FIR-radio correlation and suggests than the minimum  
energy field strength is an underestimate for luminous starbursts.   \\
\label{fig:ubup}}  
\end{figure*}  
  
\subsection{The FIR-Radio Correlation}  
\label{section:for_fir}  
  
Equation (\ref{fir_radio}) implies a linear correlation between the
FIR luminosity ($L_{\rm FIR}$) and the radio luminosity ($L_{\rm
rad}$) of galaxies.  Such a correlation is in fact observed over more
than four orders of magnitude in $L_{\rm FIR}$ (van der Kruit 1971,
1973; Helou et al.~1985; Condon 1992; Yun, Reddy, \& Condon 2001).
Equation (\ref{fir_radio}) is a consequence of the fact that if
$\tau_{\rm cool} \lesssim \tau_{\rm esc}$, then the radio flux is
essentially independent of the magnetic field strength and depends
only on the rate at which energy is supplied to the relativistic
electrons.  The linearity of the FIR-radio correlation then follows
naturally from the fact that the supernova rate is proportional to the
star formation rate (e.g., the ``calorimeter'' theory; V\"olk 1989;
see also Bressan et al.~2002).  Thus as long as $\tau_{\rm
cool} \lesssim \tau_{\rm esc}$ one can account for the observed linear
FIR-radio correlation with no fine tuning.
  
By contrast, if $\tau_{\rm cool}\gtrsim\tau_{\rm esc}$ (as argued in  
\S\ref{section:against_breaks} \& \ref{section:against_halos}) then  
significant fine tuning is {required} to explain the FIR-radio  
correlation.  Because the synchrotron cooling timescale  
(eq.~[\ref{syncool}]) varies by a factor of $\sim 10^3$ for the  
systems in Figure \ref{fig:bmin} (a factor of $\sim 60$ for the  
starbursts alone), the escape timescale would have to vary by exactly  
the same factor in order to preserve a linear FIR-radio correlation.  
We believe that this level of fine tuning argues strongly against  
$\tau_{\rm cool} \gtrsim \tau_{\rm esc}$ (and thus indirectly against  
$B \sim B_{\rm min}$).  
  
The magnitude of the radio flux from star-forming galaxies is also  
fully consistent with the hypothesis that $\tau_{\rm  
cool}\lesssim\tau_{\rm esc}$.  To demonstrate this, we note that if  
each supernova supplies an energy of $\xi 10^{48}$ ergs to cosmic ray  
electrons with a spectral index $p = 2$, and if $\tau_{\rm cool} \ll  
\tau_{\rm esc}$, then the radio luminosity should be related to the  
FIR luminosity via $\nu L_\nu \approx 10^{-5} \, \xi \,  
L_{IR}/2\ln(\gamma_{max}) \approx 2.5 \times 10^{-7} \, \xi \, L_{IR}$,  
where $\gamma_{max}$ is the maximum Lorentz factor of the accelerated  
electrons.  For comparison, the 1.4 GHz FIR-radio correlation from Yun  
et al.~(2001) is $\nu L_\nu \approx 2 \times 10^{-6} \, L_{IR}$.  This  
implies that $\xi \approx 8$ is needed to account for the FIR-radio  
correlation.  That is, $\approx 0.8 \%$ of each supernova's energy is  
radiated away as synchrotron radiation.  This is similar to standard  
estimates for the fraction of SN energy {\it supplied} to relativistic  
electrons (see, e.g., \S2.2.2 of Keshet et al.~2003 for a summary),  
which supports the hypothesis that $\tau_{\rm cool} \lesssim \tau_{\rm  
esc}$.  In particular, this argument rules out the possibility that  
$\tau_{\rm cool} \gg \tau_{\rm esc}$, since in this case the amount of  
supernova energy supplied to cosmic rays would have to be  
prohibitively large ($\xi \gg 1$).  
  
An oft-cited objection to the calorimeter model for the FIR-radio
correlation is that the nonthermal radio spectra of galaxies should
then be significantly steeper than are observed because of strong
synchrotron cooling (see \S\ref{section:against_breaks}; e.g., Condon
1992).  In \S\ref{section:for_breaks} we show that ionization and
bremsstrahlung losses can systematically flatten the radio spectra of
starbursts even when $\tau_{\rm cool}\ll\tau_{\rm esc}$, providing much
better agreement with observations.
  
Equation (\ref{taus}) implies that $\tau_{\rm syn} \gtrsim \tau_{\rm
esc}$ for $\Sigma_g \lesssim 0.005 \, (0.05)$ g cm$^{-2}$, if
$\tau_{\rm esc} \approx 3 \times 10^7 \, (10^6)$ yr. The former value
of $\tau_{\rm esc}$ is comparable to that estimated for the Milky Way
(Garcia-Munoz et al.~1977; Connell 1998).  According to the arguments
presented in this section, the calorimeter theory for the FIR-radio
correlation thus does not apply to low surface density normal
star-forming galaxies.  One might then expect a change in the
FIR-radio correlation at low $\Sigma_g$ because a smaller fraction of
the electron energy is radiated away.  Observationally, there is only
a very mild suppression in the ratio of the radio to FIR luminosities
for low-luminosity galaxies (Yun et al.~2001).  Bell (2003) has shown,
however, that this is in part because neither the radio nor the FIR
emission from low luminosity galaxies are good proxies for the total
star formation rate.  Because the focus of this paper is on starburst
galaxies, we defer a more detailed analysis of the connection between
the FIR-radio correlation in normal spirals and starbursts to a future
paper.
  
\subsubsection{Synchrotron versus Inverse Compton Cooling}  
\label{section:for_syn_ic}  
  
Regardless of the value of $\tau_{\rm cool}/\tau_{\rm esc}$, the
linearity of the FIR-radio correlation suggests that the synchrotron
cooling timescale $\tau_{\rm syn}$ (eq.~[\ref{syncool}]) must be
shorter than the IC cooling timescale $\tau_{\rm IC}$
(eq.~[\ref{iccool}]).  This in turn implies that $U_B \gtrsim U_{ph}$.
If this constraint is not satisfied, any variation in $U_B/U_{ph}$
would imply large changes in the fraction of cosmic ray electron
energy radiated via synchrotron radiation.  A linear FIR-radio
correlation would then require significant fine tuning.

In Figure \ref{fig:ubup} we plot the ratio $(B^2_{\rm
min}/8\pi)/U_{ph}$ versus gas surface density for the sample of
galaxies from Figure \ref{fig:bmin} (Tables \ref{table:spiral}-
\ref{table:extreme}).  The photon energy density is computed using
\beq U_{ph}=\frac{F}{c}=\epsilon\sds c,
\label{uph}  
\eeq   
where $F$ is the radiative flux, $\sds$ is the star formation  
rate per unit area, and we take $\epsilon = 3.8\times10^{-4}$ for  
consistency with Kennicutt (1998).  Above the dotted  
line in Figure \ref{fig:ubup}, $U_{B_{\rm min}}/U_{ph}>1$, $\tau_{\rm  
syn}<\tau_{\rm IC}$, and synchrotron dominates IC as the primary  
coolant for the relativistic electrons.  Below the dotted line  
$\tau_{\rm syn}> \tau_{\rm IC}$ and IC dominates.  
  
Figure \ref{fig:ubup} shows that for high surface density starbursts  
the minimum energy argument implies magnetic energy densities smaller  
than the photon energy density.  For the most extreme case of Arp 220  
(and ULIRGs in general), the minimum energy estimate implies  
$U_B/U_{ph} \sim 0.1$.  For most starbursts in Figure \ref{fig:ubup},  
a modest increase of $B$ by a factor of $\sim 2-3$ would lead to $U_B  
\gtrsim U_{ph}$.  This is plausibly within the uncertainties of the  
minimum energy estimate --- and is a rather minor correction on the  
scale of $B_{\rm eq}/B_{\rm min}$ in Figure \ref{fig:bmin} --- but  
does suggest that $B_{\rm min}$ is an underestimate.  
  
Equation (\ref{uph}) may significantly underestimate $U_{ph}$ for
galaxies with $\Sigma_g \gtrsim 0.5$ g cm$^{-2}$ because the ISM is
optically thick to its own FIR radiation in these systems.  In this
case, \beq U_{ph}=\frac{\tau_{IR}F}{c}\approx \tau_{IR} \epsilon\sds
c,
\label{uphtau}   
\eeq where $\tau_{IR}\approx \kappa_{IR} \Sigma_g$ is the optical
depth of the galactic disk to the IR radiation and $\kappa_{IR}
\approx 3-10$ cm$^2$ g$^{-1}$ is the temperature-dependent dust
opacity in the FIR (Thompson, Quataert, \& Murray 2005).  In
principle, this factor of $\tau_{IR}$ could increase $U_{ph}$ by a
factor of $10-100$ for ULIRGs.  If one requires that $U_B > U_{ph}$ in
order to satisfy the FIR-radio correlation, the necessary field
strength would then be $\gg B_{\rm min}$, and perhaps as large as
$\sim B_{\rm eq}$.  The applicability of equation (\ref{uphtau})
hinges, however, on the assumption that the radio emitting electrons
are co-spatial with optically thick gas at the average density of the
ISM.  If the radio emitting electrons are in a lower-than-average
density (optically thin) environment, then equation (\ref{uph}) is
still valid.
  
\subsection{Cooling Breaks}  
\label{section:for_breaks}  
  
In \S\ref{section:against_breaks} we noted that because starburst
radio spectra do not show significant evidence for synchrotron
cooling, the straightforward interpretation is that $\tau_{\rm
cool}\gtrsim\tau_{\rm esc}$ and, hence, $B\sim B_{\rm min}$.  One
difficulty with this conclusion is provided by ULIRGs such as Arp 220,
where the IC cooling time alone is $\sim 10^4$ yrs.  Unless $h
\lesssim 10$ pc, the cooling time in ULIRGs is shorter than the
advection time across the galactic disk, for reasonable values of the
advection velocity.  Given that the size of the radio emitting region
resolved by the VLA in ULIRGs is typically $\sim$ 100\,pc (Condon et
al.~1991), one might thus expect the nonthermal spectrum to be
systematically steeper in ULIRGs than in other
starbursts.\footnote{Although the radio sizes are $\sim 100$ pc, a
small scale height for ULIRGs is not ruled out: for a random velocity
$\delta v$, the scale height of a galactic disk is $h \approx \delta
v^2/2\pi G \Sigma$, which implies $h \approx 10$ pc for the observed
values of $\delta v \approx 100$ km s$^{-1}$ and $\Sigma \approx 10$ g
cm$^{-2}$ in Arp 220 (e.g., Scoville et al.~1997).}  This does not
however, appear to be the case (Condon et al.~1991; see also Fig.~19
of Downes \& Solomon 1998 for the case of Arp 220).  The lack of a
systematic difference between the radio spectral indices of ULIRGs and
other systems raises significant questions about the extent to which
the radio spectra can be easily used to constrain the electron cooling
time and thus the magnetic field strength.
  
We suggest that the radio spectra of starbursts are in fact compatible
with $\tau_{\rm cool} \ll \tau_{\rm esc}$ because of the effect of
ionization and bremsstrahlung losses on the low energy part of the
electron distribution function. The timescale for a relativistic
electron that emits synchrotron radiation at frequency $\nu$ to lose
its energy ionizing neutral hydrogen of density $n = 10 \, n_{10}$
cm$^{-3}$ is \beq \tau_{\rm ion} \approx 10^7 \, n_{10}^{-1} \,
\nughz^{1/2} \, B_{100}^{-1/2} \, {\rm yr}. \label{ion} \eeq
Relativistic electrons also lose energy to bremsstrahlung emission on
ambient neutral nuclei, on a timescale \beq \tau_{\rm brem} \approx 3
\times 10^6 \, n_{10}^{-1} \, {\rm yr}. \label{brem} \eeq Setting $B =
\eta^{1/2} B_{\rm eq}$ and $n = f \langle n \rangle = f \Sigma_g/2 h
m_p$ we find that the ratio of the ionization loss time to the
synchrotron cooling time is \beq {\tau_{\rm ion} \over \tau_{\rm syn}}
\approx 1.6 \, \nughz \, h_{100} \, f^{-1} \, \eta^{1/2}
\label{ionsync} \eeq and that the ratio of the bremsstrahlung loss time to the 
synchrotron cooling time is \beq {\tau_{\rm brem} \over \tau_{\rm syn}}
\approx 2.4 \, \nughz^{1/2} \, h_{100} \, f^{-1} \, \eta^{3/4} \,
\Sigma_g^{1/2}, \label{bremsync} \eeq where $h_{100}=h/100$ pc is the
gas scale height. Equations (\ref{ionsync}) \& (\ref{bremsync}) show
that if the cosmic ray electrons interact with gas at the mean density
of the ISM ($f \sim 1$) ionization and bremsstrahlung losses can
significantly modify the distribution function for electrons emitting
at $\sim {\rm GHz}$ frequencies.  

\begin{figure}  
\resizebox{\hsize}{!}{\includegraphics{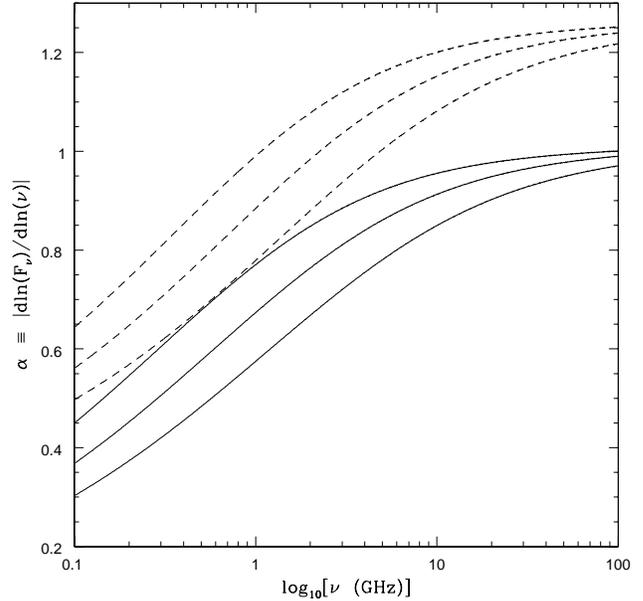}}  
\caption{Predicted radio spectral index as a function of frequency.
Electrons are injected with $n(\gamma) \propto \gamma^{-p}$, with
$p=2$ (solid lines) and $p=2.5$ (dashed lines), and lose energy
through synchrotron radiation, bremsstrahlung, and ionization
losses. Escape is neglected, which corresponds to the limit $\tau_{\rm
cool} \ll \tau_{\rm esc}$.  The calculations assume $\Sigma_g = 1$ g
cm$^{-2}$, $\eta = 1$, $h_{100} = 1$, and $f = 1, 0.5, 0.25$ (from
bottom to top). Ionization losses flatten the radio spectrum at low frequencies.
Therefore, even in the limit $\tau_{\rm cool} \ll \tau_{\rm esc}$,
spectral indices at GHz frequencies are less than the canonical value
of $\alpha = 1$ (for $p=2$) appropriate for strong synchrotron
cooling.  This provides much better agreement with the observed radio
spectra of starbursts.}
\label{fig:slope}  
\end{figure}

Under the action of ionization losses alone, low energy cosmic rays
lose energy more rapidly than high energy cosmic rays (the
bremsstrahlung loss time is independent of electron energy).
The opposite is true for synchrotron losses.  For this reason,
ionization losses can systematically decrease the spectral index from
what one would expect for a synchrotron-cooled electron distribution
($\alpha \approx 1-1.25$).  To demonstrate this explicitly, we have
calculated the steady state electron distribution function for a model
in which relativistic electrons are injected with a spectral index $p$
and lose energy via synchrotron radiation, bremsstrahlung losses, and
ionization losses.  Escape is neglected, which corresponds to the
limit $\tau_{\rm cool} \ll \tau_{\rm esc}$.  Figure \ref{fig:slope}
shows the resulting synchrotron spectral slope $\alpha$ as a function
of frequency.  The solid lines are for $p=2$ and the dashed lines are
for $p=2.5$.  The calculations shown in Figure \ref{fig:slope} assume
$\Sigma_g = 1$ g cm$^{-2}$, $\eta = 1$, $h_{100} = 1$, and $f = 1,
0.5,$ and $0.25$ (from bottom to top).  

The spectral indices in Figure \ref{fig:slope} at GHz frequencies are
compatible with the observed values even in the presence of strong
cooling.  We find similar results for reasonable variations of our
model parameters.  For example, if $p = 2$, $\eta = 0.1$, and $f =
0.25$, we find $\alpha \approx 0.6$ at 1 GHz from $\Sigma_g \approx
0.1-10$ g cm$^{-2}$ and $\alpha \approx 0.7-0.8$ at 5 GHz over the
same range of $\Sigma_g$.  Our calculations predict that the
non-thermal spectral index should steepen to $\alpha\sim1$ at
sufficiently high frequencies, but this may be difficult to detect
because of free-free emission. A second prediction of this model is
that the radio spectral index should be larger at high frequencies
than it is at low frequencies, with a transition at $\sim$ GHz
frequencies where the ionization, bremsstrahlung, and synchrotron loss
times are comparable (see eqns.~[\ref{ionsync}] \& [\ref{bremsync}]).
For example, in the model from Figure \ref{fig:slope} with $p = 2$ and
$f = 0.5$ (the middle solid line), the effective spectral index from
$1-10$ GHz is $\alpha = 0.8$, whereas from $0.1-1$ GHz it is $\alpha =
0.55$.  Using the radio data in Table \ref{table:starburst}, as well as
57.5, 151, 408, 750, and 4850 MHz data available in the NASA Extragalactic
Database (from Hales et al.~1993; Ficarra et al.~1985; Israel \&
Mahoney 1990; Becker et al.~1991; McCutcheon 1973; Waldram et
al.~1996; Heeschen \& Wade 1964; Large et al.~1981; Griffith et
al.~1995; Kuehr et al.~1981), and additional data from Oly \& Israel
(1993) (for NGC 3504), Irwin \& Saikia (2003) (for NGC 3079), and
Hummel et al.~(1991) (for NGC 891), we find that for the sample of
starburst galaxies considered in this paper the spectral index below 1
GHz is indeed typically flatter than that above 1 GHz by $\Delta
\alpha \approx 0.3$ (although there are some exceptions and scatter).
  
Although encouraging, a larger and more homogeneous sample of data is  
clearly needed to test our predictions of spectral index variations as  
a function of frequency.  In addition, low frequency radio emission  
may be strongly modified by free-free absorption in many starbursts.  
Moreover, electrons emitting at $\sim 100$ MHz in a $\sim$ mG field  
have energies of $\sim 100$ MeV.  At these energies secondary  
electrons produced by charged pion decay are important and may  
significantly modify the electron spectrum as well.

\begin{figure}  
\resizebox{\hsize}{!}{\includegraphics{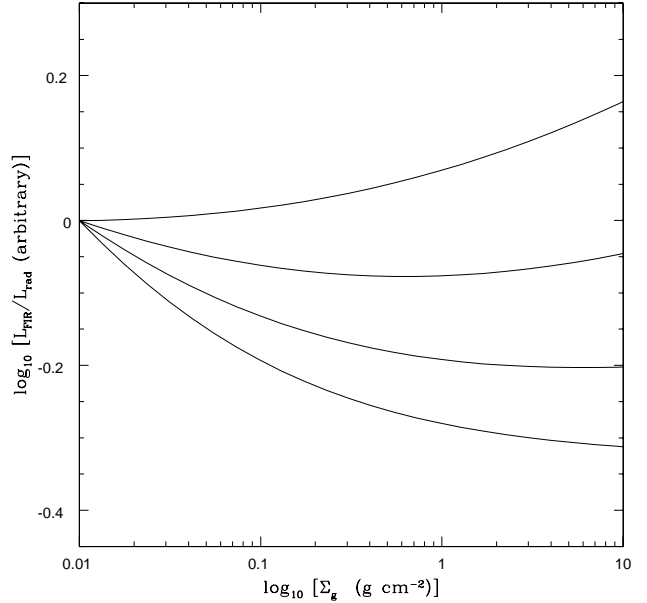}}  
\caption{Predicted ratio of the FIR to 1 GHz synchrotron luminosity as
a function of gas surface density $\Sigma_g$.  Electrons are injected
with a $p = 2$ spectrum and lose energy through synchrotron radiation
and ionization and bremsstrahlung losses; escape is neglected.  The
FIR luminosity is assumed to be proportional to the total rate at
which energy is supplied to relativistic electrons.  The calculations
assume that $f = 0.5$ ($n = 0.5 \, \langle n \rangle$) and that $B
\propto \Sigma_g^a$ with $a =$ 1, 0.9, 0.8, and 0.7 (bottom to top)
and $B = B_{\rm eq}$ at $\Sigma_g = 0.01$ g cm$^{-2}$ (see
Fig.~\ref{fig:bmin} for these scalings).  These results demonstrate
that although ionization and bremsstrahlung losses modify the radio
spectra of starbursts (Fig.~\ref{fig:slope}), they nonetheless
maintain a nearly linear FIR-radio correlation, in agreement with
observations.\\}
\label{fig:firradio}  
\end{figure}

Because the synchrotron, ionization, and bremsstrahlung loss rates all
scale differently with the gas density and magnetic field strength,
there is no guarantee that the linearity of the FIR-radio correlation
is preserved under the influence of ionization and bremsstrahlung
losses.  To assess this, Figure \ref{fig:firradio} shows model
calculations of $L_{\rm FIR}/L_{\rm rad}$ as a function of
$\Sigma_g$. The synchrotron flux at 1 GHz is calculated assuming that
electrons are injected with a $p = 2$ spectrum, and lose energy via
synchrotron, ionization, and bremmstrahlung losses (as before, escape
is neglected).  The FIR luminosity is assumed to be proportional to
the rate at which energy is supplied to relativistic electrons.
The models shown in Figure \ref{fig:firradio} assume that $f = 0.5$
($n = 0.5 \langle n \rangle)$, $h_{100} = 1$, and that $B \propto
\Sigma_g^a$ with a = $1, 0.9, 0.8,$ and $0.7$ (bottom to top; see
Fig. \ref{fig:bmin} for the $B(\Sigma_g)$ curves).  All of the assumed
models for $B(\Sigma_g)$ have $B \gg B_{\rm min}$ for luminous
starbursts.

Figure \ref{fig:firradio} demonstrates that the predicted FIR-radio
correlation is linear to better than a factor of 2 over 3 orders of
magnitude in $\Sigma_g$. The models with $a = 0.7$, $0.8$, and $0.9$ provide
particularly good agreement with the observed linearity of the
FIR-radio correlation.  Because galaxies of a given $\Sigma_g$ can
have a range of FIR luminosities (depending on galaxy size), the
systematic trends in Figure \ref{fig:firradio} may show up in part as
scatter in the canonical plots of $L_{\rm FIR}/L_{\rm rad}$
vs. $L_{\rm FIR}$; the $\lesssim$ factor of 2 variation in Figure
\ref{fig:firradio} is consistent with the observed scatter in the
FIR-radio correlation.  The approximate constancy of $L_{\rm
FIR}/L_{\rm rad}$ in Figure \ref{fig:firradio} arises because if $f
\sim 1$ and $B \gg B_{\rm min}$, then $\tau_{\rm syn}/(\tau_{\rm
ion}^{-1} + \tau^{-1}_{\rm brem})^{-1}$ is relatively constant over
most of the parameter space relevant to observed starbursts.  

In \S \ref{section:for_fir} we estimated that $\approx 0.8 \%$ of the
energy of each supernova must be supplied to relativistic electrons to
account for the magnitude of the $\approx 1.4$ GHz radio flux from
starburst galaxies, taking into account synchrotron radiation alone
and assuming electrons are injected with a $p = 2$ spectrum.  If
ionization and bremsstrahlung losses are important in starbursts, as
argued in this section, then the required energy injection rate
increases to $\approx 1.6 \%$ of the supernova energy.  In addition, if
the electron injection spectrum is $p = 2.2$, then the required energy
injection increases by an additional factor $\approx 2$.

In contrast to the results of Figure \ref{fig:firradio}, in a model in
which $f = 0.5$ and $B \propto \Sigma_g^{0.4}$, which corresponds to
the observed scaling of $B_{\rm min}(\Sigma_g)$ in Figure
\ref{fig:bmin}, the ratio of the FIR to radio luminosities increases
by a factor of $\approx 10$ from $\Sigma_g = 0.01$ g cm$^{-2}$ to
$\Sigma_g = 10$ g cm$^{-2}$ because of the increasing suppression of
the radio flux by bremsstrahlung losses.  Thus field strengths
$\approx B_{\rm min}$ are only consistent with the observed radio
emission from starbursts if cosmic-ray electrons interact with gas at
much less than the mean density of the ISM ($f \ll 1$).

To summarize this subsection, if $B \gg B_{\rm min}$ and if cosmic-ray
electrons interact with gas at approximately the mean density of the
ISM, ionization and bremsstrahlung losses will systematically flatten
the nonthermal spectra of starburst galaxies (Fig. \ref{fig:slope})
and yet maintain the linearity of the FIR-radio correlation
(Fig. \ref{fig:firradio}).
This eliminates one of the strongest arguments against $\tau_{\rm
cool} \ll \tau_{\rm esc}$ and argues for $B \gg B_{\rm min}$ in
starbursts.
  
\subsection{Pion Production}  
\label{section:for_gamma}  
  
An independent test of the density of the ISM in which the cosmic rays
reside can be provided by gamma-ray emission from neutral pion decay.
The timescale for cosmic ray protons to lose their energy to pion
production is $t_{\rm pion} \sim 10^8 \, (n/1 \, {\rm cm^{-3}})^{-1}$
yr.  If $n \sim \langle n \rangle$, this is significantly shorter than
$\tau_{\rm esc}$ in starbursts (see eq.~[\ref{tau_esc}]) and so pion
production will lead to a substantial $\gamma$-ray luminosity.
Assuming that $10\%$ of the energy of each supernova is supplied to
relativistic protons with a spectral index $p = 2$, and that $\approx
1/3$ of this energy goes into $\gamma$-rays rather than neutrinos or
pairs, we estimate a $\gamma$-ray luminosity from neutral pion decay
in starbursts of $\nu L_\nu \approx 3 \times 10^{-4}
L_{IR}/\ln(\gamma_{max}) \approx 2 \times 10^{-5} \, L_{IR}$, where
$\gamma_{max}$ is the maximum Lorentz factor of the accelerated
protons and the estimate is only applicable above $\approx 100$ MeV
(see Torres 2004 for more detailed calculations).  This predicted flux
is just below the EGRET upper limits for systems such as M82 and Arp
220 (Cillis et al.~2005), but should be detectable with {\it GLAST}.
Because the timescales for electron ionization and bremsstrahlung
losses and proton pion losses are quite similar, detection of
gamma-ray emission at this level would also confirm the importance of
ionization and bremsstrahlung losses for the observed radio spectra of
starbursts.

Relativistic protons lose $\approx 2/3$ of their energy to the
production of charged pions, which subsequently decay into high energy
neutrinos and electron/positron pairs; the pairs receive $\approx 1/4$
of the pion energy.  Thus if $\sim 10 \%$ of the energy of each
supernova is supplied to relativistic protons, and if $t_{\rm pion}
\ll t_{\rm esc}$, then $\sim 1.5 \%$ of the energy of each
supernova goes into secondary electrons and positrons.  This is in
good agreement with the energy injection rate estimated in
\S\ref{section:for_fir} and \S\ref{section:for_breaks} to account for
the radio flux from starbursts.  This suggests that secondary
electrons may dominate the radio emission in starbursts
(see also Rengarajan 2005).  If correct,
it remains to be understood why normal spirals and starbursts lie on
approximately the same FIR-radio correlation even though secondaries
do not appear to be important for electrons emitting at $\sim$ GHz
frequencies in normal spirals (Strong et al.~2004).
As discussed in \S\ref{section:for_fir} variations in 
$\tau_{\rm syn}/\tau_{\rm esc}$ might also be expected to modify 
the FIR-radio correlation for normal spirals relative to starbursts.
  
\subsection{Inverse Compton Emission}  
\label{section:for_ic}  
  
IC upscattering of infrared photons by cosmic-ray electrons can  
provide an independent constraint on $U_B/U_{ph}$ because the ratio of  
the IC power to the radio power is given by $L_{IC}/L_{\rm rad}  
\approx U_{ph}/U_B$. If we use equation (\ref{uph}) to estimate  
$U_{ph}$, then $B \sim B_{\rm eq}$ in starbursts would lead to  
negligible IC power because it would imply $U_B/U_{ph} \gg 1$.  
However, if the infrared optical depth is larger than unity, equation  
(\ref{uphtau}) may instead be applicable, in which case $B \sim B_{\rm  
eq}$ implies $U_B/U_{ph} \sim 1$.  
   
IC emission has long been argued to contribute significantly to the  
X-ray emission in a variety of systems, including M82 (e.g., Hargrave  
1974; Moran \& Lehnert 1997) and NGC 3256 (e.g., Moran, Lehnert, \&  
Helfand 1999).  In both of these examples, however, much of the hard  
X-ray emission has been resolved into point sources with {\it  
Chandra}, although IC emission may still contribute to the hard X-rays  
from the very central nucleus of the starburst (e.g., Lira et  
al.~2002; Strickland et al.~2002; Strickland \& Heckman, private  
communication).  More detailed tests of this hypothesis would be very  
worthwhile given the possibility of providing an independent  
constraint on the magnetic field strength in starbursts.  
  
\subsection{Synchrotron Halos}  
\label{section:for_halos}  
  
As an alternative to the standard interpretation (\S
\ref{section:against_halos}) that extended synchrotron halos are due
to relativistic electrons generated in the disk and advected into the
halo (before cooling), we suggest that the halos are instead due to
electrons accelerated {\it in situ} in the galactic wind, by either
direct electron acceleration in shocks or via charged pion production.
We note that in several simulations of galactic winds (e.g.,
Strickland \& Stevens 2000), most of the X-ray emission is due to {\it
in situ} shocks, which are also likely to be sites of particle
acceleration.  Since the total kinetic energy flux in a galactic wind
is comparable to the energy produced by supernovae in the starburst,
only a small fraction of the wind energy must be converted into
relativistic particles {\it in situ} to account for the extended
synchrotron emission observed.  In this interpretation, which clearly
requires additional work, $B \gg B_{\rm min}$ would be fully
consistent with extended synchrotron emission.  Observed spectral
differences between the galactic disk and halo in M82 and other
systems may arise from spatial variations in $\tau_{\rm syn}/\tau_{\rm
esc}$, $\tau_{\rm IC}/\tau_{\rm esc}$, $\tau_{\rm brem}/\tau_{\rm syn}$, 
and $\tau_{\rm ion}/\tau_{\rm syn}$.
  
\section{Discussion \& Conclusions}  
\label{section:discussion}  
  
Based on the analysis of the preceding sections, it is worthwhile  
contrasting the following two scenarios for the radio emission from  
starbursts:  
  
\noindent {\it $B \sim B_{\rm min}$:} For the minimum energy magnetic
field estimate to be applicable, the electrons must escape rapidly in
a galactic wind with $\tau_{\rm cool} \gtrsim \tau_{\rm esc}$.  The
rapid escape implies that there is no cooling break at GHz
frequencies, which is consistent with the observed nonthermal spectra
of $\alpha \approx 0.75$.  Rapid escape is plausible given that hot
outflowing galactic winds are observed from starbursts and that the
cosmic rays are likely generated in the same supernova shocks that
generate the wind.  This interpretation of magnetic fields in
starbursts --- the standard one --- has two major drawbacks: (1) it is
difficult to see how $\tau_{\rm cool} \gtrsim \tau_{\rm esc}$ is
applicable in ULIRGs where cooling times from IC alone are $\sim 10^4$
yr and (2) it is very difficult to account for the FIR-radio
correlation with $\tau_{\rm cool} \gtrsim \tau_{\rm esc}$ because
variations in $\tau_{\rm cool}$ or $\tau_{\rm esc}$ should lead to
variations in the fraction of the electron energy radiated away via
synchrotron radiation.  Either several coincidences or a complex
feedback loop are then required to account for the linearity of the
FIR-radio correlation.  The latter would be somewhat surprising given
that $B_{\rm min}$ itself implies that magnetic fields and cosmic rays
are dynamically weak compared to gravity (Fig.~\ref{fig:bmin}).
  
\noindent {\it $B \gg B_{\rm min}$:} An alternative possibility is
that magnetic fields in starbursts are much stronger than is suggested
by the minimum energy estimate.  Rapid cooling of relativistic
electrons in starbursts will invalidate the minimum energy estimate if
$\tau_{\rm cool} \ll \tau_{\rm esc}$.  Magnetic fields in starbursts
could then in principle be as large as $\sim B_{\rm eq}$, which is
$\sim 10$ times larger than $B_{\rm min}$ for typical starbursts
(Fig.~\ref{fig:bmin}).
This model naturally accounts for the linearity of the FIR-radio
correlation because in the limit $\tau_{\rm cool} \lesssim \tau_{\rm
esc}$ the radio flux is nearly independent of the magnetic field
strength (e.g., V\"olk 1989). In addition, the observed radio
luminosities from star-forming galaxies are comparable to the
estimated rate at which energy is supplied to relativistic electrons
in supernova shocks (\S\ref{section:for_fir}).  This supports the
hypothesis that $\tau_{\rm cool} \lesssim \tau_{\rm esc}$.  Independent
support for this hypothesis is provided by the fact that the observed
scaling $B_{\rm min} \propto \Sigma_g^{2/5}$ for the minimum energy
field in starbursts (Fig. \ref{fig:bmin}) follows directly from the
Schmidt Law for star formation when $\tau_{\rm cool}\lesssim\tau_{\rm
esc}$ (eq. [\ref{bminscale}]).
  
A standard objection to $\tau_{\rm cool} \lesssim \tau_{\rm esc}$ is
that the typical nonthermal spectral indices of starbursts ($\alpha
\approx 0.75$) do not show the expected steepening due to strong
cooling ($\alpha \approx 1$).  We have argued, however, that if $B \gg
B_{\rm min}$ and if cosmic ray electrons interact with gas at
approximately the mean density of the ISM, then ionization and
bremsstrahlung losses flatten the radio spectra of starbursts at $\sim
{\rm GHz}$ frequencies and reconcile $\tau_{\rm cool} \lesssim
\tau_{\rm esc}$ with the observed spectra (Fig.~\ref{fig:slope}).  An
important part of this argument is the realization that the ionization
and bremsstrahlung loss times are similar to the synchrotron cooling
time in all starburst galaxies for cosmic ray electrons emitting at
GHz frequencies.  Thus, ionization and bremsstrahlung losses can
modify the nonthermal spectra of starburst galaxies and yet maintain
the linearity of the FIR-radio correlation (Fig. \ref{fig:firradio}).
  
This interpretation of magnetic fields in starbursts --- that $B \gg
B_{\rm min}$ --- has two potential drawbacks: (1) Given that galactic
winds efficiently remove mass and metals from galaxies, it is unclear
whether the cosmic rays actually interact with the bulk of the ISM in
starbursts, which is required for ionization losses to be significant.
(2) The nonthermal synchrotron halos observed in several systems are
typically interpreted as cosmic-ray electrons advected out with a
galactic wind, with synchrotron cooling only becoming important in the
halo (e.g., Seaquist \& Odegard 1991).  If this interpretation is
correct, it requires $B \sim B_{\rm min}$
(\S\ref{section:against_halos}).  We have suggested, however, that
extended synchrotron emission may be due to particles accelerated {\it
in situ} in galactic winds (\S\ref{section:for_halos}), in which case
the observed halos cannot be readily used to constrain the magnetic
field strength in galactic disks.
  
We believe that the arguments presented in this paper strongly favor
$\tau_{\rm cool} \lesssim \tau_{\rm esc}$ for the relativistic
electrons in starbursts.  In this limit, the minimum energy argument
underestimates the true magnetic field strength.  The observed radio
flux is, however, nearly independent of the magnetic field and depends
primarily on the rate at which energy is supplied to relativistic
electrons.  Thus direct constraints on the true field strength are
difficult to come by.  By themselves, these arguments do not imply
that $B\sim B_{\rm eq}$, only that $\tau_{\rm cool} \lesssim \tau_{\rm
esc}$ and $B \gtrsim B_{\rm min}$.  Our strongest argument that
magnetic fields are likely to be $\gg B_{\rm min}$ derives from the
ratio of the ionization and bremsstrahlung loss times to the
synchrotron cooling time: if cosmic ray electrons interact with gas
having roughly the mean density of the ISM, magnetic fields {\it must}
be $\gg B_{\rm min}$ in starbursts in order for ionization and
bremsstrahlung losses to not completely dominate over synchrotron
losses (which would be inconsistent with the FIR-radio correlation and
would require that an unreasonably large fraction of the supernova
energy be supplied to cosmic ray electrons in order to account for the
observed radio fluxes from starbursts; \S\ref{section:for_breaks}).
  
Our conjecture is thus that the magnetic field is significantly larger  
than $\sim B_{\rm min}$ (the canonical estimate) in many starbursts.
Confirming this prediction would have significant implications for
understanding the physics of star formation in starbursts, the effects
of magnetic fields on the dynamics of galactic winds, and the role of
magnetic stresses in driving gas to smaller radii to fuel a central
active galactic nucleus.  Towards this end, we briefly summarize
several observations that should help test the predictions of our
model: (1) If ionization and bremsstrahlung losses are important, the
radio spectra of starbursts should be flatter below $\sim 1$ GHz than
they are above $\sim 1$ GHz (\S\ref{section:for_breaks} and
Fig. \ref{fig:slope}).  (2) If cosmic rays interact with gas at about
the mean density of the ISM, as is required for ionization and
bremsstrahlung losses to be important, there should be an appreciable
gamma-ray flux from starbursts due to neutral pion production
(\S\ref{section:for_gamma}).  We estimate $\nu L_\nu \approx 2 \times
10^{-5} L_{IR}$ above $\approx 100$ MeV.  This prediction is testable
with {\it GLAST}. (3) Zeeman measurements of starbursts can directly
probe the magnetic field strength in the dense phase of the ISM, and
distinguish between $B \sim B_{\rm min}$ and $B \gg B_{\rm min}$.
Existing upper limits in 4 ULIRGs from Killeen al.~(1996) are above
$B_{\rm min}$, but below the equipartition field
(\S\ref{section:against_zeeman}).  (4) If $B \gg B_{\rm min}$, IC
emission is unlikely to contribute significantly to the X-ray emission
in starbursts (\S\ref{section:for_ic}).
  
It is interesting to compare our results on magnetic fields in
starbursts with the inferred correlation between magnetic field
strength, column density, and mass density in Galactic molecular
clouds (e.g., Troland \& Heiles 1986; Crutcher 1999).  Using Zeeman
splitting to determine the magnetic field strength directly, Crutcher
(1999) finds that over the range $0.01 \, {\rm g \, cm^{-2}} \lesssim
\Sigma_{\rm H_2} \lesssim 2 \, {\rm g \, cm^{-2}} \lesssim$ and
$2.7\lesssim \log_{10}[n_{\rm H_2}\,({\rm cm}^{-3})]\lesssim6.8$, that
$B\propto n_{\rm H_2}^{1/2}$ and $B\propto \Sigma_{\rm H_2}$. These
findings imply both that that the Alfv\'en speed $v_A \sim {\rm
constant} \sim 10$ km s$^{-1}$ and that equipartition obtains.
Observations of polarization in H$_2$O masers in dense star-forming
regions provide evidence that these trends extend to yet higher
densities ($10^8-10^{10}$ cm$^{-3}$; Sarma et al.~2002; Vlemmings et
al.~2005).  These results suggest that the magnetic field is always
comparable to the total pressure in Galactic molecular clouds, with
$B\approx B_{\rm eq}$.
  
The sample of Galactic molecular clouds reviewed by Crutcher (1999)  
covers precisely the same surface densities represented by starbursts  
in our Figure \ref{fig:bmin}.  It is thus particularly striking that  
the minimum energy estimate implies $B \ll B_{\rm eq}$ in starbursts,  
while direct Zeeman measurements of local molecular clouds with the  
same $\Sigma_g$ imply $B \sim B_{\rm eq}$.  Our suggestion is that  
this difference arises in part from the inapplicability of the minimum  
energy estimate in starbursts.  
  
\acknowledgments
  
We are grateful to Rainer Beck, Tim Heckman, and Jim Condon
for a close reading of the text and for a number of very useful
comments.  We also thank Alberto Bolatto, Bruce Partridge, and Carl
Heiles for valuable conversations.  E.Q.~is supported in part by NASA
grant ATP05-54, an Alfred P.~Sloan Fellowship, and the David and
Lucile Packard Foundation.  E.W.~thanks the Miller Foundation for
supporting his visit to UC Berkeley and the Institute for Advanced
Study.  N.M.~is supported in part by a Canadian Research Chair in
Astrophysics.  This research made extensive use of the NASA
Extragalactic Database.  C.L.M.~is supported by the David and Lucile
Packard foundation and Alfred P.~Sloan Fellowship.


\newpage  
  
\begin{table}  
\begin{scriptsize}  
\begin{center}  
\caption{Properties of Normal Star-Forming Galaxies \label{table:spiral}}  
  
\begin{tabular}{lccccccccccc}  
\hline \hline  
  
\\  
  
\multicolumn{1}{c}{Object Name} &  
\multicolumn{1}{c}{$z$\tablenotemark{a}} &   
\multicolumn{1}{c}{$D$\tablenotemark{b}} &   
\multicolumn{1}{c}{$\log_{10}[\Sigma_g]$\tablenotemark{c,d}} &  
\multicolumn{1}{c}{$\log_{10}[\sds]$\tablenotemark{e,d}} &   
\multicolumn{1}{c}{$S_{1.4\,{\rm GHz}}$\tablenotemark{f}}&  
\multicolumn{1}{c}{$\log_{10}[L_{\rm rad}]$\tablenotemark{g}}&  
\multicolumn{1}{c}{Angular Diameter\tablenotemark{d}} &   
\multicolumn{1}{c}{Radius}&  
\multicolumn{1}{c}{$B_{\rm min}$\tablenotemark{h}}&  
\multicolumn{1}{c}{$B_{\rm eq}$\tablenotemark{i}}&  
\multicolumn{1}{c}{Radio Data}\\  
  
\multicolumn{1}{c}{} &  
\multicolumn{1}{c}{} &   
\multicolumn{1}{c}{(Mpc)} &   
\multicolumn{1}{c}{$({\rm g\,cm^{-2}})$} &  
\multicolumn{1}{c}{$(M_\odot{\rm \,yr^{-1}\,kpc^{-2}})$} &  
\multicolumn{1}{c}{(mJy)} &   
\multicolumn{1}{c}{(W Hz$^{-1}$)} &   
\multicolumn{1}{c}{($\arcsec$)} &   
\multicolumn{1}{c}{(kpc)} &  
\multicolumn{1}{c}{($\mu$G)}&  
\multicolumn{1}{c}{($\mu$G)}&  
\multicolumn{1}{c}{References} \\  
  
\\  
\hline  
\\  
  
NGC 224....................&  \nodata &      0.9 &    $-$3.00 &    $-$3.13 &   8400   &    20.89 &     9912 &     20.9 &      3.3 &      2.3 &   1\\   
NGC 598....................&  \nodata &      0.9 &    $-$2.65 &    $-$2.47 &   3300   &    20.51 &     3354 &      7.3 &      4.6 &      5.1 &   1\\   
NGC 628....................&   0.0022 &      9.3 &    $-$2.75 &    $-$2.18 &    180   &    21.27 &      612 &     13.8 &      5.3 &      4.1 &   1\\   
NGC 772....................&   0.0082 &     34.6 &    $-$2.74 &    $-$2.84 &     71.4 &    22.01 &      432 &     36.3 &      5.0 &      4.2 &   1\\   
NGC 925....................&  \nodata &      9.3 &    $-$2.77 &    $-$2.44 &     46.0 &    20.68 &      534 &     12.0 &      3.9 &      3.9 &   1\\   
NGC 1569.................. &  \nodata &      2.2 &    $-$2.35 &    $-$0.80 &    411   &    20.40 &      150 &      0.82 &     15.1 &     10.2 &  1\\   
NGC 2336.................. &   0.0074 &     31.1 &    $-$2.77 &    $-$1.92 &     17.7 &    21.31 &      414 &     31.2 &      3.4 &      3.9 &   1\\   
NGC 2403.................. &  \nodata &      3.3 &    $-$2.80 &    $-$2.15 &    330   &    20.64 &      948 &      7.6 &      4.9 &      3.6 &   1\\   
NGC 2841.................. &   0.0021 &      9.0 &    $-$2.71 &    $-$2.99 &     83.8 &    20.91 &      408 &      8.9 &      5.4 &      4.5 &   1\\   
NGC 2903.................. &  \nodata &      6.1 &    $-$2.82 &    $-$2.31 &    407   &    21.26 &      642 &      9.5 &      6.5 &      3.5 &   1\\   
NGC 2976.................. &  \nodata &      3.4 &    $-$2.70 &    $-$1.66 &     50.8 &    19.85 &      294 &      2.4 &      5.6 &      4.6 &   1\\   
NGC 3031.................. &  \nodata &      3.3 &    $-$2.83 &    $-$2.50 &    380   &    20.70 &     1332 &     10.6 &      4.2 &      3.4 &   1\\   
NGC 3310.................. &   0.0033 &     13.9 &    $-$2.54 &    $-$1.14 &    383   &    21.95 &      210 &      7.1 &     12.2 &      6.6 &   1\\   
NGC 3338.................. &   0.0043 &     18.3 &    $-$2.87 &    $-$2.56 &     28.1 &    21.06 &      354 &     15.7 &      4.3 &      3.1 &   1\\   
NGC 3368.................. &   0.0030 &     12.6 &    $-$2.75 &    $-$2.55 &     30.0 &    20.76 &      390 &     12.0 &      4.1 &      4.1 &   1\\   
NGC 3486.................. &   0.0023 &      9.6 &    $-$2.80 &    $-$2.46 &     59.4 &    20.82 &      426 &      9.9 &      4.8 &      3.6 &   1\\   
NGC 3521.................. &   0.0027 &     11.3 &    $-$2.46 &    $-$1.91 &    357   &    21.74 &      486 &     13.4 &      7.4 &      7.9 &   1\\   
NGC 3631.................. &   0.0039 &     16.3 &    $-$2.52 &    $-$1.73 &     80.8 &    21.41 &      276 &     10.9 &      6.7 &      6.9 &   1\\   
NGC 3675.................. &   0.0026 &     10.8 &    $-$2.69 &    $-$2.01 &     43.7 &    20.79 &      354 &      9.3 &      4.9 &      4.7 &   1\\   
NGC 3726.................. &   0.0028 &     12.0 &    $-$2.62 &    $-$2.28 &     27.4 &    20.68 &      324 &      9.4 &      4.5 &      5.5 &   1\\   
NGC 3893.................. &   0.0032 &     13.7 &    $-$2.62 &    $-$1.96 &    134   &    21.48 &      234 &      7.8 &      8.5 &      5.5 &   1\\   
NGC 3938.................. &   0.0027 &     11.4 &    $-$2.48 &    $-$2.11 &     61.7 &    20.99 &      318 &      8.8 &      5.7 &      7.6 &   1\\   
NGC 4178.................. &   0.0013 &      5.3 &    $-$2.55 &    $-$2.27 &     23.3 &    19.90 &      240 &      3.1 &      5.1 &      6.4 &   1\\   
NGC 4254.................. &   0.0080 &     33.9 &    $-$2.29 &    $-$1.70 &    422   &    22.77 &      312 &     25.7 &     10.0 &     11.7 &   1\\   
NGC 4258.................. &   0.0015 &      6.3 &    $-$3.09 &    $-$2.36 &    790   &    21.58 &      906 &     13.9 &      6.5 &      1.9 &   1\\   
NGC 4294.................. &   0.0012 &      5.0 &    $-$2.66 &    $-$1.87 &     24.9 &    19.88 &      150 &      1.8 &      6.8 &      5.0 &   2\\   
NGC 4303.................. &   0.0052 &     22.1 &    $-$2.47 &    $-$1.74 &    416   &    22.39 &      354 &     19.0 &      9.2 &      7.7 &   1\\   
NGC 4321.................. &   0.0053 &     22.4 &    $-$2.54 &    $-$2.07 &    340   &    22.31 &      408 &     22.1 &      8.1 &      6.6 &   1\\   
NGC 4394.................. &   0.0031 &     13.0 &    $-$3.04 &    $-$2.88 &      0.7 &    19.15 &      234 &      7.4 &      1.9 &      2.1 &   1\\   
NGC 4402.................. &   0.0008 &      3.3 &    $-$2.60 &    $-$2.80 &     52.2 &    19.85 &      186 &      1.5 &      7.4 &      5.7 &   3\\   
NGC 4501.................. &   0.0076 &     32.1 &    $-$2.59 &    $-$2.21 &    278   &    22.54 &      360 &     28.0 &      8.2 &      5.9 &   1\\   
NGC 4519.................. &   0.0041 &     17.2 &    $-$2.69 &    $-$1.98 &      9.1 &    20.51 &      186 &      7.8 &      4.5 &      4.7 &   2\\   
NGC 4535.................. &   0.0065 &     27.6 &    $-$2.67 &    $-$2.38 &     64.5 &    21.77 &      378 &     25.3 &      5.2 &      4.9 &   1\\   
NGC 4548.................. &   0.0016 &      6.8 &    $-$2.99 &    $-$2.52 &      4.6 &    19.42 &      306 &      5.1 &      2.8 &      2.3 &   1\\   
NGC 4569.................. &  \nodata &     18.0 &    $-$3.07 &    $-$2.78 &     83.4 &    21.51 &      474 &     20.7 &      4.9 &      1.9 &   1\\   
NGC 4571.................. &  \nodata &     15.0 &    $-$2.85 &    $-$2.56 &      4.2 &    20.06 &      222 &      8.1 &      3.2 &      3.2 &   1\\   
NGC 4579.................. &   0.0051 &     21.4 &    $-$2.87 &    $-$2.32 &    103   &    21.76 &      306 &     15.9 &      6.7 &      3.1 &   1\\   
NGC 4647.................. &   0.0047 &     19.9 &    $-$2.64 &    $-$2.22 &     35.0 &    21.23 &      204 &      9.9 &      6.2 &      5.2 &   3\\   
NGC 4651.................. &   0.0028 &     12.0 &    $-$2.62 &    $-$1.98 &     24.0 &    20.62 &      210 &      6.1 &      5.5 &      5.5 &   1\\   
NGC 4654.................. &   0.0035 &     14.6 &    $-$2.58 &    $-$2.06 &    117   &    21.48 &      258 &      9.1 &      7.7 &      6.0 &   1\\   
NGC 4689.................. &   0.0054 &     22.8 &    $-$2.74 &    $-$2.38 &      9.6 &    20.78 &      234 &     12.9 &      4.0 &      4.2 &   1\\   
NGC 4698.................. &   0.0033 &     14.1 &    $-$3.43 &    $-$3.55 &      0.6 &    19.16 &      222 &      7.6 &      1.9 &      0.8 &   1\\   
NGC 4713.................. &   0.0022 &      9.2 &    $-$2.64 &    $-$1.53 &     37.8 &    20.59 &      156 &      3.5 &      7.4 &      5.2 &   2\\   
NGC 4736.................. &   0.0010 &      4.4 &    $-$3.03 &    $-$2.22 &    254   &    20.78 &      630 &      6.8 &      5.8 &      2.1 &   1\\   
NGC 4826.................. &   0.0014 &      5.8 &    $-$3.01 &    $-$2.47 &    103   &    20.61 &      480 &      6.7 &      5.2 &      2.2 &   1\\   
NGC 5033.................. &   0.0029 &     12.4 &    $-$2.75 &    $-$2.64 &    178   &    21.52 &      546 &     16.4 &      5.7 &      4.1 &   1\\   
NGC 5055.................. &   0.0017 &      7.1 &    $-$2.51 &    $-$2.32 &    390   &    21.38 &      660 &     11.4 &      6.4 &      7.1 &   1\\   
NGC 5194.................. &   0.0015 &      6.5 &    $-$2.21 &    $-$1.78 &   1490   &    21.88 &      600 &      9.4 &      9.8 &     14.1 &   1\\   
NGC 5236.................. &   0.0017 &      7.3 &    $-$1.98 &    $-$1.41 &   2445   &    22.19 &      660 &     11.6 &     10.7 &     23.9 &   1\\   
NGC 5457.................. &  \nodata &      8.0 &    $-$2.59 &    $-$2.46 &    750   &    21.77 &     1614 &     31.4 &      4.6 &      5.9 &   1\\   
NGC 6217.................. &   0.0045 &     19.2 &    $-$2.39 &    $-$1.91 &     78.4 &    21.54 &      180 &      8.4 &      8.4 &      9.3 &   1\\   
NGC 6503.................. &  \nodata &      5.2 &    $-$2.79 &    $-$2.08 &     36.5 &    20.08 &      294 &      3.7 &      5.1 &      3.7 &   1\\   
NGC 6643.................. &   0.0049 &     20.9 &    $-$2.57 &    $-$1.81 &     92.2 &    21.69 &      204 &     10.3 &      8.2 &      6.2 &   1\\   
NGC 6946.................. &  \nodata &      3.0 &    $-$2.38 &    $-$1.88 &   1395   &    21.17 &      642 &      4.6 &      9.3 &      9.5 &   1\\   
NGC 7331.................. &   0.0027 &     11.6 &    $-$2.60 &    $-$2.33 &    373   &    21.78 &      510 &     14.3 &      7.3 &      5.7 &   1\\   
  
\\  
  
\hline  
\hline  
\end{tabular}  
\end{center}  
\end{scriptsize}  
  
\tablenotetext{a}{Redshifts from NED. $^{\rm b}$Distance. In most cases, distances are computed from $z$   
with $H=71$ km s$^{-1}$ Mpc$^{-1}$.  In cases where redshifts are not a reliable distance indicator, $D$ is taken from the literature.  
$^{\rm c}$Gas surface density.  
$^{\rm d}$Data from Kennicutt (1998).  
$^{\rm e}$Star formation rate per unit area.   
$^{\rm f}$Radio flux density listed at 1.4 and 1.49 GHz.   
$^{\rm g}$Radio luminosity computed using $L_{\rm rad}=4\pi D^2 S_{1.4\,{\rm GHz}}$.   
$^{\rm h}$$B_{\rm min}$ calculated using equation (\ref{bmin}), with $\nu_{\rm GHz}=1.49$, $V=2\pi R^2 h_{\rm rad}$,   
and assuming $h_{\rm rad}=500$\,pc.  
$^{\rm i}$$B_{\rm eq}$ calculated using equation (\ref{bequipartition}).}  
  
\tablerefs{(1) Condon (1987); (2) Condon, Yin, \& Burstein (1987); (3) Condon et al.~(1990).}  
  
\end{table}


\begin{table}  
\begin{scriptsize}  
\begin{center}  
\caption{Properties of Starburst Galaxies \label{table:starburst}}  
  
\begin{tabular}{lccccccccccc}  
\hline \hline  
  
\\  
  
\multicolumn{1}{c}{Object Name} &  
\multicolumn{1}{c}{$z$\tablenotemark{a}} &   
\multicolumn{1}{c}{$D$\tablenotemark{b}} &   
\multicolumn{1}{c}{$\log_{10}[\Sigma_g]$\tablenotemark{c,d}} &  
\multicolumn{1}{c}{$\log_{10}[\sds]$\tablenotemark{e,d}} &   
\multicolumn{1}{c}{$S_{1.4\,{\rm GHz}}$\tablenotemark{f}}&  
\multicolumn{1}{c}{$\log_{10}[L_{\rm rad}]$\tablenotemark{g}}&  
\multicolumn{1}{c}{Angular Diameter\tablenotemark{d}} &   
\multicolumn{1}{c}{Radius}&  
\multicolumn{1}{c}{$B_{\rm min}$\tablenotemark{h}}&  
\multicolumn{1}{c}{$B_{\rm eq}$\tablenotemark{i}}&  
\multicolumn{1}{c}{Radio Data}\\  
  
\multicolumn{1}{c}{} &  
\multicolumn{1}{c}{} &   
\multicolumn{1}{c}{(Mpc)} &   
\multicolumn{1}{c}{$({\rm g\,cm^{-2}})$} &  
\multicolumn{1}{c}{$(M_\odot{\rm \,yr^{-1}\,kpc^{-2}})$} &  
\multicolumn{1}{c}{(mJy)} &   
\multicolumn{1}{c}{(W Hz$^{-1}$)} &   
\multicolumn{1}{c}{($\arcsec$)} &   
\multicolumn{1}{c}{(kpc)} &  
\multicolumn{1}{c}{(mG)}&  
\multicolumn{1}{c}{(mG)}&  
\multicolumn{1}{c}{References} \\  
  
\\  
\hline  
\\  
NGC 253....................     &  \nodata &      3.5 &    $-$0.33 &     1.24 &   5594 &     21.93 &      24 &     0.21 &     0.14 &     1.1  &  1 \\   
NGC 520....................     &   0.0074 &     31.2 &       0.13 &     1.32 &    167 &     22.29 &       5 &     0.38 &     0.13 &     3.1  &  2 \\   
NGC 660....................     &   0.0028 &     12.0 &    $-$1.08 &     0.06 &    387 &     21.83 &      31 &     0.90 &     0.06 &     0.19 &  2 \\   
NGC 828....................     &   0.0179 &     75.7 &    $-$0.02 &     1.10 &    108 &     22.87 &       5 &     0.92 &     0.11 &     2.2  &  3 \\   
NGC 891....................     &   0.0018 &      7.4 &    $-$1.07 &  $-$0.58 &    701 &     21.67 &      35 &     0.63 &     0.06 &     0.19 &  3 \\   
NGC 1097..................      &   0.0043 &     18.0 &    $-$1.01 &  $-$0.20 &    415 &     22.21 &      35 &     1.52 &     0.05 &     0.22 &  3 \\   
NGC 1614..................      &   0.0159 &     67.3 &       0.04 &     1.79 &    123 &     22.83 &       4 &     0.65 &     0.13 &     2.5  &  2 \\   
NGC 1808..................      &   0.0034 &     14.2 &    $-$1.03 &     0.08 &    519 &     22.10 &      30 &     1.03 &     0.06 &     0.21 &  1 \\   
NGC 2146..................      &   0.0030 &     12.6 &    $-$0.85 &     0.84 &   1087 &     22.32 &      17 &     0.52 &     0.11 &     0.32 &  1 \\   
NGC 2623..................      &   0.0185 &     78.0 &    $-$0.81 &     1.00 &   98.5 &     22.86 &       8 &     1.51 &     0.08 &     0.35 &  2 \\   
NGC 3034 (M82)......            &  \nodata &      3.3 &    $-$0.16 &     1.48 &   7657 &     22.00 &      29 &     0.23 &     0.14 &     1.6  &  2 \\   
NGC 3079..................      &   0.0038 &     15.9 &       0.57 &     1.63 &    849 &     22.41 &       5 &     0.19 &     0.20 &     8.5  &  2 \\   
NGC 3351..................      &   0.0026 &     11.0 &    $-$0.85 &     0.24 &   47.8 &     20.84 &      14 &     0.37 &     0.05 &     0.32 &  2 \\   
NGC 3504..................      &   0.0051 &     21.7 &    $-$0.78 &     0.11 &    265 &     22.18 &      16 &     0.84 &     0.08 &     0.38 &  2 \\   
NGC 3627..................      &   0.0024 &     10.2 &    $-$1.40 &  $-$0.77 &    458 &     21.76 &      39 &     0.97 &     0.05 &     0.09 &  2 \\   
NGC 3690..................      &   0.0100 &     42.2 &    $-$1.40 &  $-$0.10 &    658 &     23.15 &      24 &     2.46 &     0.08 &     0.09 &  2 \\   
NGC 6240..................      &   0.0243 &    102.7 &       0.43 &     1.87 &    392 &     23.70 &       3 &     0.75 &     0.22 &      6.2 &  3 \\   
NGC 7552..................      &   0.0053 &     22.3 &    $-$1.30 &     0.16 &    276 &     22.22 &      22 &     1.19 &     0.06 &     0.11 &  1 \\   
IC 342........................  &   0.0010 &      4.4 &    $-$1.62 &  $-$0.41 &   2250 &     21.71 &      67 &     0.71 &     0.06 &     0.05 &  1 \\   
IC 694........................  &   0.0104 &     43.8 &       0.42 &     2.40 &    153 &     22.55 &     2.6 &     0.28 &     0.18 &     6.0  &  4 \\   
IC 883 (Arp 193).......         &   0.0233 &     98.5 &       0.27 &     1.54 &    101 &     23.07 &     3.4 &     0.81 &     0.14 &     4.26 &  2 \\   
IC 1623......................   &   0.0204 &     86.1 &       0.13 &     1.67 &    221 &     23.30 &     3.6 &     0.75 &     0.17 &     3.1  &  2 \\   
Maffei 2.....................   &  \nodata &      3.0 &    $-$1.22 &  $-$0.27 &    286 &     20.48 &      40 &     0.29 &     0.05 &     0.14 &  5 \\   
Arp 55........................  &   0.0393 &    165.9 &    $-$0.95 &     0.32 &     33 &     23.04 &       8 &     3.22 &     0.06 &     0.26 &  6 \\   
Arp 220......................   &   0.0181 &     76.6 &       1.08 &     2.98 &    324 &     23.36 &       2 &     0.37 &     0.26 &     27.5 &  2 \\   
IRAS\,10173+0828.....           &   0.0480 &    202.8 &    $-$1.27 &     0.48 &    9.9 &     22.69 &       7 &     3.44 &     0.05 &     0.12 &  6 \\   
IRAS\,17208$-$0014.....         &   0.0425 &    179.4 &       0.41 &     2.01 &     73 &     23.45 &       3 &     1.30 &     0.14 &     5.9  &  6 \\   
VII Zw 31..................	&   0.0537 &    226.8 &    $-$0.57 &     0.82 &   41.6 &     23.41 &       5 &     2.75 &     0.09 &     0.62 &  3 \\   
Zw 049.057...............	&   0.0332 &    140.1 &       0.22 &     1.77 &   46.4 &     23.04 &       3 &     1.02 &     0.12 &     3.8  &  2 \\   
\hline  
\hline  
\end{tabular}  
\end{center}  
\end{scriptsize}  
  
\tablenotetext{a}{Redshifts from NED. $^{\rm b}$Distance. In most cases, distances computed with $H=71$ km s$^{-1}$ Mpc$^{-1}$.   
In some cases, distances are taken from the literature instead of using the Hubble Law.  
$^{\rm c}$Gas surface density.  
$^{\rm d}$Data taken from Kennicutt (1998).  
$^{\rm e}$Star formation rate per unit area.  
$^{\rm f}$Radio flux density listed at 1.40, 1.425, or 1.49 GHz.   
$^{\rm g}$Radio luminosity computed using $L_{\rm rad}=4\pi D^2 S_{1.4\,{\rm GHz}}$.   
$^{\rm h}$$B_{\rm min}$ calculated using equation (\ref{bmin}), with $\nu_{\rm GHz}=1.49$,  
$V=2\pi R^2 h_{\rm rad}$, and assuming $h_{\rm rad}=100$\,pc.  
$^{\rm i}$$B_{\rm eq}$ calculated using equation (\ref{bequipartition}).}  
  
\tablerefs{(1) Condon (1987); (2) Condon et al.~(1990); (3) Condon et al.~(1996);   
(4) Condon \& Broderick (1988); (5) Condon et al.~(1998); (6) Yun, Reddy, \& Condon (2001).}  
  
\end{table}


\begin{table}  
\begin{scriptsize}  
\begin{center}  
\caption{Properties of ``Extreme'' Starburst Galaxies \label{table:extreme}}  
  
\begin{tabular}{lccccccccccc}  
\hline \hline  
  
\\  
  
\multicolumn{1}{c}{Object Name} &  
\multicolumn{1}{c}{$z$\tablenotemark{a}} &   
\multicolumn{1}{c}{$D$\tablenotemark{b}} &   
\multicolumn{1}{c}{$\log_{10}[\Sigma_g]$\tablenotemark{c}} &  
\multicolumn{1}{c}{$\log_{10}[\sds]$\tablenotemark{d}} &   
\multicolumn{1}{c}{$S_{8.44\,{\rm GHz}}$\tablenotemark{e}}&  
\multicolumn{1}{c}{$\log_{10}[L_{\rm rad}]$\tablenotemark{f}}&  
\multicolumn{1}{c}{Angular Diameter\tablenotemark{c}} &   
\multicolumn{1}{c}{Radius}&  
\multicolumn{1}{c}{$B_{\rm min}$\tablenotemark{g}}&  
\multicolumn{1}{c}{$B_{\rm eq}$\tablenotemark{h}}&  
\multicolumn{1}{c}{Radio Data}\\  
  
\multicolumn{1}{c}{} &  
\multicolumn{1}{c}{} &   
\multicolumn{1}{c}{(Mpc)} &   
\multicolumn{1}{c}{$({\rm g\,cm^{-2}})$} &  
\multicolumn{1}{c}{$(M_\odot{\rm \,yr^{-1}\,kpc^{-2}})$} &  
\multicolumn{1}{c}{(mJy)} &   
\multicolumn{1}{c}{(W Hz$^{-1}$)} &   
\multicolumn{1}{c}{($\arcsec$)} &   
\multicolumn{1}{c}{(kpc)} &  
\multicolumn{1}{c}{(mG)}&  
\multicolumn{1}{c}{(mG)}&  
\multicolumn{1}{c}{References} \\  
\\  
\hline  
\\  
IC 883 (Arp 193).......      &   0.0233 &     98.5 &     0.25 &     2.68 &     34.9 &    22.61 &     0.69 &     0.16 &     0.33 &     4.06 & 1 \\   
Mrk 273..................... &   0.0378 &    159.6 &     0.66 &     3.36 &     43.5 &    23.13 &     0.35 &     0.14 &     0.51 &    10.45 & 1 \\   
Arp 220 (west)...........    &   0.0181 &     76.6 &     0.94 &     3.55 &     72.2 &    22.71 &     0.40 &     0.07 &     0.55 &    19.91 & 1 \\   
Arp 220 (east)............   &   0.0181 &     76.6 &     0.78 &     2.96 &     60.7 &    22.63 &     0.66 &     0.12 &     0.39 &    13.77 & 1 \\   
\\  
  
\hline  
\hline  
\end{tabular}  
\end{center}  
\end{scriptsize}  
  
\tablenotetext{a}{Redshifts from NED. $^{\rm b}$Distances computed with $H=71$ km s$^{-1}$ Mpc$^{-1}$.   
$^{\rm c}$Gas surface density from Downes \& Solomon (1998), Tables 3, 4, and 12. Values adjusted for assumed $H_0$.   
$^{\rm d}$Star formation rate per unit area computed using $\sds=L_{\rm IR}(<R)/(\pi R^2\epsilon c^2)$  
with $\epsilon=3.8\times10^{-4}$ (see Downes \& Solomon 1998 Table 12).  
$^{\rm e}$Radio flux density at 8.44 GHz.   
$^{\rm f}$Radio luminosity computed using $L_{\rm rad}=4\pi D^2 S_{8.44\,{\rm GHz}}$.   
$^{\rm g}$$B_{\rm min}$ calculated using equation (\ref{bmin}), with $\nu_{\rm GHz}=8.44$,  
$V=2\pi R^2 h_{\rm rad}$, and assuming $h_{\rm rad}=100$\,pc.  
$^{\rm h}$$B_{\rm eq}$ calculated using equation (\ref{bequipartition}).}  
  
\tablerefs{(1) Condon et al.~(1991).}  
  
\end{table}

\end{document}